%% file: main.tex
\documentclass[10pt, conference]{IEEEtran}
\usepackage{cite}
\usepackage{amsmath,amssymb,amsfonts}
\usepackage{algorithmic}
\usepackage{graphicx}
\usepackage{textcomp}
\usepackage{xcolor}

\input{content/variables}

\input{content/packages}

\title{\paperTitle}

\author{\IEEEauthorblockN{Pramod Chunduri}
\IEEEauthorblockA{\textit{Georgia Institute of Technology}\\
pramodc@gatech.edu}
\and
\IEEEauthorblockN{Yao Lu}
\IEEEauthorblockA{\textit{National University of Singapore}\\
luyao@comp.nus.edu.sg}
\and
\IEEEauthorblockN{Joy Arulraj}
\IEEEauthorblockA{\textit{Georgia Institute of Technology}\\
arulraj@gatech.edu}
}
\begin{document}

% \author{{Pramod Chunduri}}
% \affiliation{%
%   \institution{Georgia Institute of Technology}
% }
% \email{pramodc@gatech.edu}

% \author{Yao Lu}
% \affiliation{%
%   \institution{Microsoft Research}
% }
% \email{luyao@microsoft.com}

% \author{Joy Arulraj}
% \affiliation{%
%   \institution{Georgia Institute of Technology}
% }
% \email{arulraj@gatech.edu}
% 

%\newcommand{\revision}[1]{\color{webblue}{#1}\unskip\color{black}}
%\newcommand{\change}[1]{\color{red}{#1}\unskip\color{black}}
%\newcommand{\takeout}[1]{\color{green}{#1}\unskip\color{black}}
%\newcommand{\change}[1]{#1}
%\newcommand{\takeout}[1]{#1}

\maketitle

\input{content/abstract}


% \pagestyle{\vldbpagestyle}
% \begingroup\small\noindent\raggedright\textbf{PVLDB Reference Format:}\\
% \vldbauthors. \vldbtitle. PVLDB, \vldbvolume(\vldbissue): \vldbpages, \vldbyear.\\
% \href{https://doi.org/\vldbdoi}{doi:\vldbdoi}
% \endgroup
% \begingroup
% \renewcommand\thefootnote{}\footnote{\noindent
% This work is licensed under the Creative Commons BY-NC-ND 4.0 International License. Visit \url{https://creativecommons.org/licenses/by-nc-nd/4.0/} to view a copy of this license. For any use beyond those covered by this license, obtain permission by emailing \href{mailto:info@vldb.org}{info@vldb.org}. Copyright is held by the owner/author(s). Publication rights licensed to the VLDB Endowment. \\
% \raggedright Proceedings of the VLDB Endowment, Vol. \vldbvolume, No. \vldbissue\ %
% ISSN 2150-8097. \\
% \href{https://doi.org/\vldbdoi}{doi:\vldbdoi} \\
% }\addtocounter{footnote}{-1}\endgroup

% \ifdefempty{\vldbavailabilityurl}{}{
% \vspace{.3cm}
% \begingroup\small\noindent\raggedright\textbf{PVLDB Artifact Availability:}\\
% The source code, data, and/or other artifacts have been made available at \url{\vldbavailabilityurl}.
% \endgroup

\input{content/introduction}

\input{content/background}

% \input{content/motivation}
% \input{content/overview}

\input{content/solution}
\input{content/evaluation}

\input{content/relatedwork}

\input{content/conclusion}

%\input{content/acknowledgement}

%% ==================================================================
%% BIBLIOGRAPHY
%% ==================================================================

% \newpage
%\clearpage
% \bibliographystyle{abbrv}
% \bibliographystyle{unsrt}
\newpage{
\bibliographystyle{IEEEtran}
\raggedright
%\small
\bibliography{main}
}

\end{document}

%% file: content/variables.tex
\newcommand{\sys}{\mbox{\textsc{Tracer}}\xspace}

\newcommand{\paperTitle}{\sys: Efficient Object Re-Identification in Networked Cameras through Adaptive Query Processing}

\newcommand{\naive}{\textsc{Na\"ive}\xspace}
\newcommand{\pp}{\textsc{PP}\xspace}
\newcommand{\spatula}{\textsc{Spatula}\xspace}
\newcommand{\dfs}{\textsc{Graph-Search}\xspace}
\newcommand{\oracle}{\textsc{Oracle}\xspace}
\newcommand{\ngram}{\textsc{N-Gram}\xspace}
\newcommand{\mle}{\textsc{MLE}\xspace}
\newcommand{\rnn}{\textsc{RNN}\xspace}
\newcommand{\reid}{\mbox{\textsc{Re-id}}\xspace}

\newcommand{\porto}{\mbox{\textsc{Porto}}\xspace}
\newcommand{\beijing}{\mbox{\textsc{Beijing}}\xspace}
\newcommand{\townfive}{\mbox{\textsc{Town05}}\xspace}
\newcommand{\townseven}{\mbox{\textsc{Town07}}\xspace}

%% file: content/packages.tex
\usepackage{anyfontsize}
\usepackage[hyphens]{url}

\definecolor{webblue}{rgb}{0,0,.7}
\definecolor{webgreen}{rgb}{0,.5,0}
\definecolor{webbrown}{rgb}{.6,0,0}
\usepackage{xstring}
\usepackage{hyperref}
\hypersetup{
  colorlinks=true,
  linkcolor=webbrown,
  citecolor=green,
  urlcolor=webblue,
}

\usepackage{amsmath,amsopn,amssymb}
\usepackage[cal=boondoxo]{mathalfa}
\usepackage{subcaption}
\usepackage{endnotes,microtype,xspace,graphicx,fancyvrb,multirow}
\usepackage{supertabular,booktabs}
\usepackage{array,underscore, relsize}
\usepackage{fancyhdr}
\usepackage{enumitem}
\usepackage{balance}
\usepackage{booktabs}
\usepackage{pifont}
\usepackage{listings}
\usepackage{paralist}

\usepackage{multirow}
\usepackage[scaled]{beramono}
\usepackage{tabularx}
\usepackage{multirow}

\makeatletter
\newcommand\BeraMonottfamily{%
  \def\fvm@Scale{0.85}% scales the font down
  \fontfamily{fvm}\selectfont% selects the Bera Mono font
}
\makeatother

\definecolor{mymauve}{rgb}{0.58,0,0.82}

\lstdefinestyle{SQLStyle}{
  language=SQL,
  basicstyle={\small\ttfamily},
  breaklines=true,
  frame=none,
  numbers=none,
  keepspaces=true,
  captionpos=b,
  stringstyle=\color{mymauve},
  keywordstyle=\color{blue},
  commentstyle=\color{dkgreen},
}

\lstdefinestyle{ScriStyle}{
language=SQL,
basicstyle=\BeraMonottfamily\footnotesize, 
keywordstyle=\color{smtcolor}\bfseries,
morekeywords={and, or, not},
aboveskip = 0.05in,
belowskip = 0.05in,
literate = {-}{-}1, % <------ trick!
}

\usepackage{aliascnt}

\usepackage[linesnumbered,ruled,vlined, noend]{algorithm2e}
\newlength\mylen

\SetCommentSty{mycommfont}
\usepackage{setspace}
\usepackage{enumitem}
\usepackage[color=notecolor]{todonotes} % provides the \todo{} command

\SetAlFnt{\small}
\SetAlCapFnt{\small}
\SetAlCapNameFnt{\small}

\usepackage{semantic}

\usepackage{stmaryrd}
\usepackage{ltablex}
\usepackage{mathtools}
\usepackage{adjustbox}
\usepackage{fixltx2e}
\usepackage[group-separator={,}]{siunitx}
\usepackage[capitalize,noabbrev,nameinlink]{cleveref}

\crefname{lstlisting}{listing}{listings}
\Crefname{lstlisting}{Listing}{Listings}

\usepackage{chngcntr}
\counterwithout{equation}{section}

\newtheorem{problem}{Problem}

\newcommand{\hide}[1]{}

\newcommand{\PP}[1]{
\vspace{2px}
\noindent{\bf\textsc{#1}.}\xspace
}

\newcommand{\PPP}[1]{
\vspace{0.05in}
\noindent{\textit{\IfEndWith{#1}{.}{#1}{#1.}}}
}

\newcommand{\squishitemize}{
 \begin{list}{$\bullet$}
  { \setlength{\itemsep}{0pt}
     \setlength{\parsep}{0pt}
     \setlength{\topsep}{0pt}
     \setlength{\partopsep}{0pt}
     \setlength{\leftmargin}{1.95em}
     \setlength{\labelwidth}{1.5em}
     \setlength{\labelsep}{0.5em} } }

\newcounter{Lcount}
\newcommand{\squishlist}{
    \begin{list}{\arabic{Lcount}. }
   { \usecounter{Lcount}
        \setlength{\itemsep}{0pt}
        \setlength{\parsep}{3pt}
        \setlength{\topsep}{0pt}
        \setlength{\partopsep}{0pt}
        \setlength{\leftmargin}{2em}
        \setlength{\labelwidth}{1.5em}
        \setlength{\labelsep}{0.5em} } }

\newcommand{\squishend}{\end{list}}

\newcommand{\bit}{\begin{compactitem}}
\newcommand{\eit}{\end{compactitem}}
\newcommand{\ben}{\begin{compactenum}}
\newcommand{\een}{\end{compactenum}}

\newcommand{\eg}{\textit{e}.\textit{g}.,\xspace}
\newcommand{\ie}{\textit{i}.\textit{e}.,\xspace}
\newcommand{\etc}{\textit{e}.\textit{t}.\textit{c}.}

\definecolor{dkgreen}{rgb}{0,0.6,0}

\newcommand{\cmark}{\ding{51}}

\def\Snospace~{\S{}}

%% file: content/abstract.tex
\begin{abstract}
%
% \notice{Go over~\cite{beck1993how} for tips on writing a good abstract.}
%
% \notice{What is the problem and why is it important?}
%
Efficiently re-identifying and tracking objects across a network of cameras is crucial for applications like traffic surveillance.
Spatula is the state-of-the-art video database management system (VDBMS) for processing \reid queries.
However, it suffers from two limitations.
Its spatio-temporal filtering scheme has limited accuracy on large camera networks due to localized camera history.
It is not suitable for critical video analytics applications that require high recall due to lack of support for adaptive query processing.

In this paper, we present \sys, a novel VDBMS for efficiently processing \reid queries using an adaptive query processing framework.
\sys selects the optimal camera to process at each time step by training a recurrent network to model long-term historical correlations.
To accelerate queries under a high recall constraint, \sys incorporates a probabilistic adaptive search model that processes camera feeds in incremental search windows and dynamically updates the sampling probabilities using an exploration-exploitation strategy.
To address the paucity of benchmarks for the \reid task due to privacy concerns, we present a novel synthetic benchmark for generating multi-camera \reid datasets based on real-world traffic distribution.
%
%This benchmark is suitable for evaluating video DBMSs on other vision queries as well.
%
Our evaluation shows that \sys outperforms the state-of-the-art cross-camera analytics system by 3.9$\times$ on average across diverse datasets.
%

% The source code, data, and/or other artifacts have been made available at: \href{https://anonymous.4open.science/r/tracer-6B51/}{https://anonymous.4open.science/r/tracer-6B51/}.
% Spotting anomalies in graphs is an important topic.
%
% \notice{Why does prior work not solve it?} 
% Prior work on finding anomalies suffer from two limitations.
%
% First, \ldots.
%
% Second, \ldots.
%

% \notice{What is your solution? Why is it novel?} 
%
% We present a better method, called \sys, to find anomalies.
%
% List the benefits of the approach - NOT the low level details of how you do it!
%
% \sys has the following properties:
% (a) {\em scale}, being linear on the input size
% (b) {\em effective}, spotting $90\%$ of the anomalies in real data
% (c) {\em automatic}, requiring no user-defined parameters.
%
% \notice{What are the implications of your solution?}
%
% Give some performance numbers.
%
% Experiments on 3~GB of real data from epinions.com illustrate the
% benefits of \sys./
%
% It is 5$\times$ faster than \baseline.
%
\end{abstract}

%% file: content/introduction.tex
\section{Introduction}
\label{sec:introduction}

% \squishitemize 
% \item \PC{2nd pass main todos:}
% \item \PC{Add a picture in the introduction with different images of the same vehicle at different intersections and maybe combine it with the Reid pipeline image}
% \item \PC{Explore exploit narrative for probabilistic search in the context of eddies/adaptive query processing}
% \item \PC{Long-term cross-camera correlations}
% \item \PC{Remove temporal filtering -- only to begin the search, present in all baselines, mention it in a paragraph to say it is used in all methods}
% \item \PC{Explicitly mention reuse potential -- for queries within the same time window}
% \item \PC{Explicitly mention distributed execution possibilities - potential for improvement}
% \item \PC{Reid benchmark comparison with syntheticle and VisualRoad}
% \squishend

% \notice{Go over~\cite{widom2006tips} for tips on writing technical
% papers.}

% \squishlist
% \item \notice{What is the problem?}
% \item \notice{Why is it interesting and important?}
% \item \notice{Why is it hard? (E.g., why do naive approaches fail?)}
% \item \notice{Why hasn't it been solved before? (Or, what's wrong with previous proposed solutions? How does mine differ?)}
% \item \notice{What are the key components of my approach and results? Also include any specific limitations.}
% \squishend

% \notice{THE PROBLEM AND ITS IMPORTANCE -- What is the problem? Why is it
% interesting and important?}
%
Advances in computer vision over the last decade have led to increased interest in the development of Video Database Management Systems (VDBMSs) that automatically analyze videos at scale~\cite{pp, blazeit, eva, vocal, viva}.
%
% Modern VDBMSs have shown great promise in a variety of applications, including surveillance, traffic monitoring, and urban planning.
%
These systems primarily focus on -- 
(1) \textit{object queries} - retrieving general instances of objects categories (\eg cars, animals)~\cite{noscope, pp, blazeit},
(2) \textit{action queries} - retrieving action segments (\eg left turn of a car, pedestrian crossing from left to right)~\cite{zeus, equivocal, videomonitoring},
(3) \textit{object track queries} - tracking a specific object within a single camera~\cite{miris, otif}.

\begin{figure}[t!]
    % \begin{adjustbox}{width=\columnwidth,center}
    \centering
    \includegraphics[width=0.55\columnwidth, height=0.35\columnwidth]{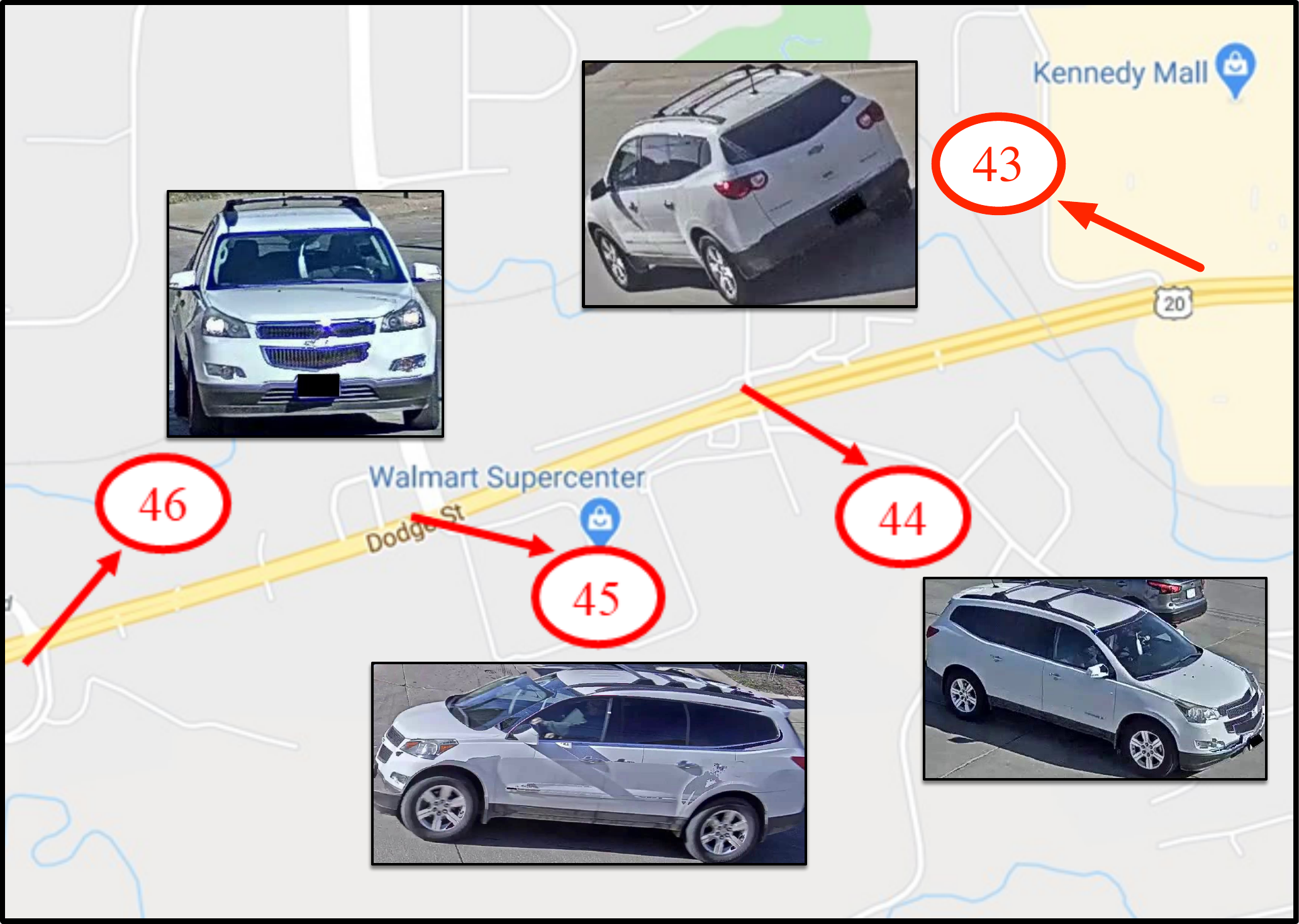}
    % \end{adjustbox}
    \caption{Object Re-identification over a Camera Network.}
    \label{fig:intro-example}
\end{figure}

In this paper, we focus on another important computer vision task -- \textit{re-identifying} and tracking an object across a network of cameras~\cite{multicamerasurvey, cityflow}, as shown in \cref{fig:intro-example}.
This \textit{object re-identification} query has many real-world applications.
For instance, a traffic monitoring expert might be interested in retrieving the trajectory of a suspicious vehicle from extensive video footage captured within a city environment.
An insurance professional may seek to retrospectively examine particular vehicle trajectories to verify instances of insurance fraud.

%
% As shown in~\cref{fig:multi-camera-tracking-example}, given a source image, the system must re-identify and track a specific vehicle across a city-scale camera network.
%

\PP{Re-identification task}
The task of multi-camera object re-identification (\reid) requires identifying all instances of an object across a camera network using a single query image of the object~\cite{cityflow}.
A \naive pipeline for processing a \textit{\reid query} involves running - 
(1) an object detector (\eg YOLOv5~\cite{yolov5}) on individual frames of all the cameras,
(2) a re-identification model~(\eg fine-grained ResNet backbones~\cite{reidbaseline}) on all the detected objects to extract fine-grained features, and 
(3) similarity matching with the source image feature vector.

As we are only interested in \textit{spotting} the object in each camera, the VDBMS terminates the search within a camera once the object is found.
The high computational cost of object detection and subsequent feature extraction makes it impractical to implement this pipeline across the entire camera network.
The pipeline runs at a low frame rate of 1~fps on a CPU and 10~fps on a server-grade GPU.

Modern VDBMSs propose proxy model-based filtering techniques to quickly filter out frames that do not contain the object category of interest (\eg car)~\cite{noscope, pp, blazeit}.
However, these techniques only work well when the video dataset exhibits a low occupancy rate~\cite{blazeit}.
City-scale traffic scenarios rarely exhibit this behavior due to high traffic density, resulting in a limited performance gain.

\begin{figure}
    \centering
    \begin{subfigure}[b]{0.49\columnwidth}
        \centering
        \includegraphics[width=\linewidth]{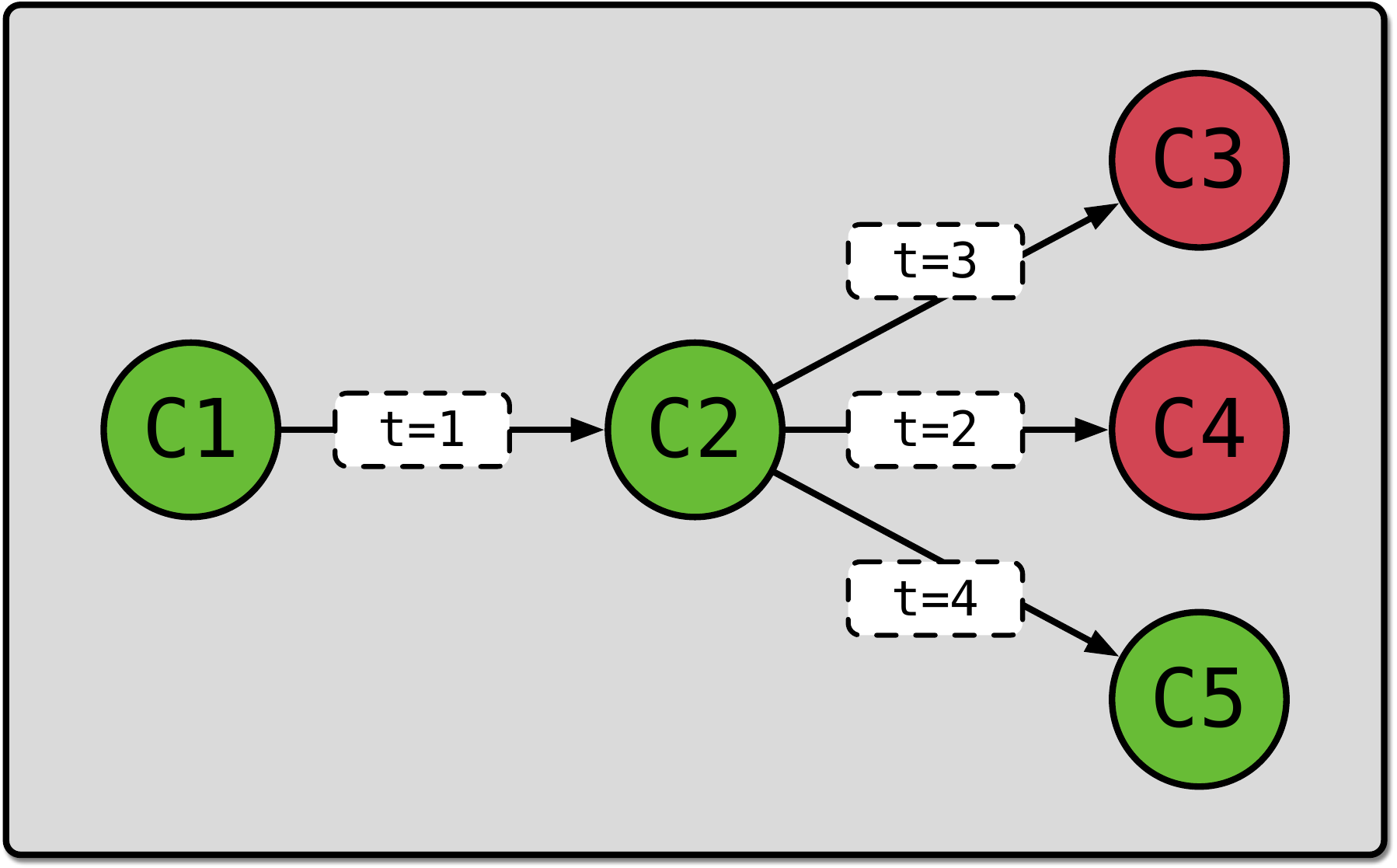}
           \caption{
                \dfs traverses the camera network in random order.
            }
           \label{fig:intro-graph-search}
    \end{subfigure}
    \hfill
    \begin{subfigure}[b]{0.49\columnwidth}
        \centering
        \includegraphics[width=\linewidth]{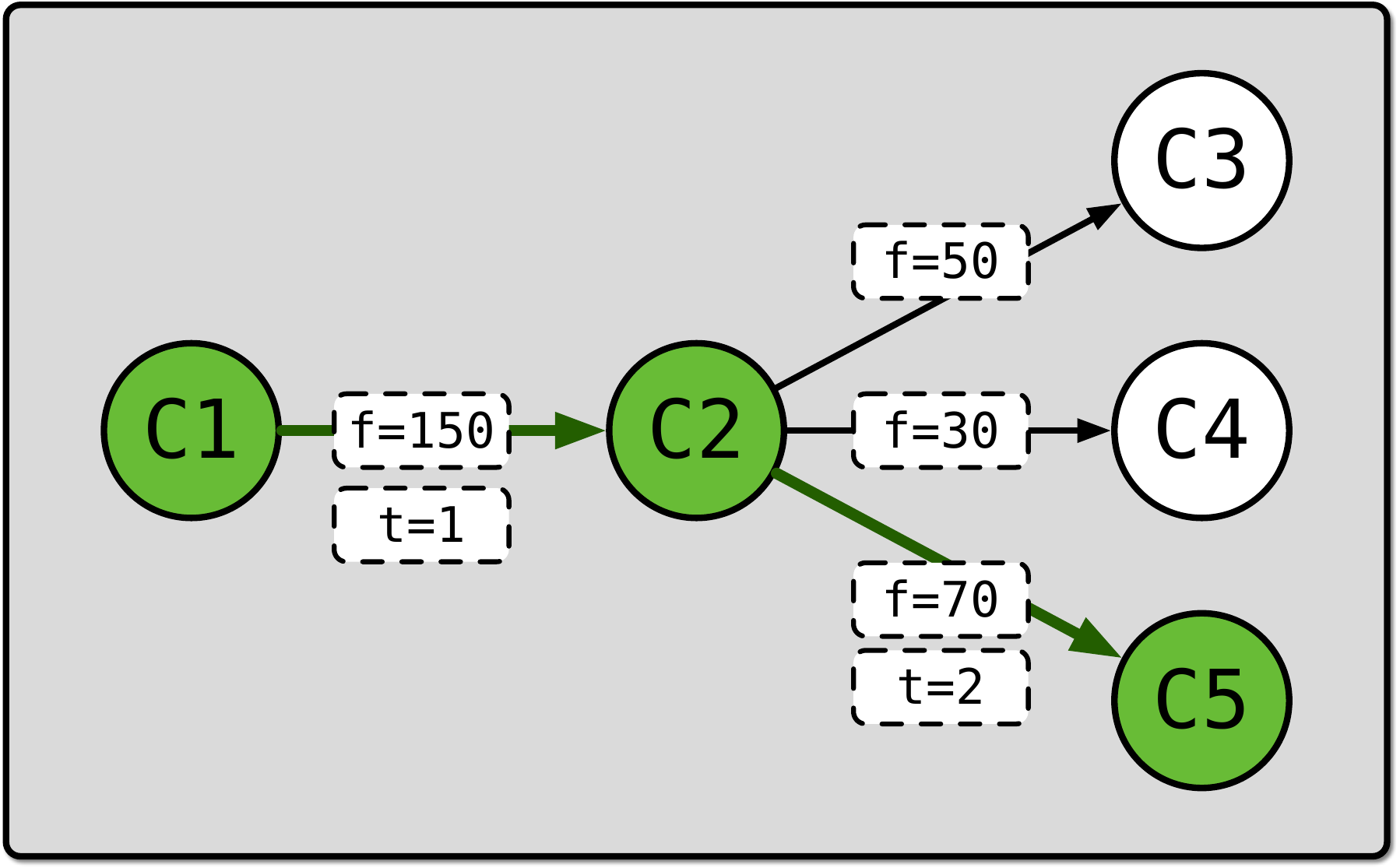}
        \caption{
            \spatula uses spatial correlations to predict the most probable camera.
        }
        \label{fig:intro-spatula}
    \end{subfigure}
    \caption{ 
    	Baseline approaches for processing \reid queries.
     }
     \label{fig:intro-baselines}
\end{figure}

\PP{Baseline \#1: Graph Search}
The camera network can be represented as a graph wherein cameras are depicted as nodes, and edges connect cameras that are directly accessible from one another, such as through adjacent roads.
Given a query object with a source camera and a time stamp, the \reid query reduces to finding the path traversed by the object through the \textit{camera graph}.
As shown in~\cref{fig:intro-graph-search}, a baseline algorithm for finding the object trajectory consists of iteratively traversing the neighboring cameras of each camera wherein the object is spotted.
Similar to canonical graph traversal algorithms,
this algorithm explores the neighboring cameras in random order.
We refer to this algorithm as \dfs.

\PP{Baseline \#2: Spatula}
Spatula~\cite{spatula} is the state-of-the-art (SoTA) system for cross-camera video analytics.
%
%\PC{Spatula is a 2020 paper - but follow-up papers focus mainly on \textit{edge} aspects, while Spatula also works in offline settings. Would calling it SoTA be fine?}
% JOY: Yes
%
As shown in~\cref{fig:intro-spatula}, Spatula uses spatio-temporal filtering on top of the camera graph to process \reid queries.
More specifically, \spatula uses \textit{localized camera history} at each camera to predict the order of the most probable neighboring cameras and the timestamp to contain the object.
In the example shown in~\cref{fig:intro-spatula}, \spatula predicts C5 as the next camera from C2 since 70 objects in the past traversed from C2 to C5 while only 50 and 30 objects traversed to C3 and C4, respectively.
%
% Similarly, it predicts that the object can be found 100 seconds after the current timestamp based on the average time taken by the object to travel from C1 to C3.
%
This way, \spatula generates a dense spatial cross-camera correlation matrix to accelerate \reid queries.
While \spatula improves upon \dfs by using a localized camera history, both techniques suffer from three limitations that constrain their runtime performance in practice.

% \PP{para 2 - Re-identification task}
% \squishitemize
% \item Re-identification (ReID) is the task of identifying a person in a video\dots
% \item What is the pipeline for re-identification
% \item What tools are necessary for reid
% \squishend

% \begin{figure*}
%     \centering
%     \includegraphics[width=\textwidth]{figures/overview/example.pdf}
%        \caption{
%             %
%             \Description{}{Graph search system for multi-camera \reid.}
%         }
%        \label{fig:spatial-filtering}
% \end{figure*}

% \notice{LIMITATIONS OF PRIOR WORK -- 
% Why is it hard? (E.g., why do naive approaches fail?)
% Why hasn't it been solved before? (Or, what's wrong with previous proposed
% solutions?)}

\PP{Challenges}
Efficiently processing \reid queries requires solving three main challenges:

\PPP{1. Large Search Space}
The first challenge is that the search space 
for the query object in a \reid query is quite large.
A 24-hour video dataset obtained from a city camera network, comprising 1000 cameras capturing footage at 10 frames per second, consists of approximately 0.9 billion frames.
Thus, the goal is to considerably reduce the number of frames to examine for each \reid query.
\dfs reduces the search space by leveraging the underlying graph structure of the camera network and avoiding irrelevant cameras (\eg far apart cameras).
However, the random exploration strategy employed in \dfs still results in a significant computational cost when the camera network exhibits high connectivity (degree), as the system may need to process multiple camera feeds before locating the correct camera.
\spatula reduces the search space by predicting a ranked ordering of the most probable (next camera, next frame) pair that contains the object using a localized camera history.
However, the spatio-temporal filtering model in \spatula is not always accurate as it relies on localized camera history to compute correlations between adjacent cameras.
This often results in the processing of incorrect neighboring cameras and start frames.
%

% \squishitemize
% \item Identifying the exact same object instead of a similar object
% \item Identifying the same object across multiple cameras with different orientations and lighting conditions
% \item Simple filtering used in frame level analysis cannot be used
% \squishend

\PPP{2. Recall-oriented processing}
The second challenge is that the target applications of \reid queries often require a high recall.
Consider a traffic monitoring expert tracking an Amber alert.
Nearly all cameras containing the suspect vehicle must be retrieved.
To achieve high recall, the system needs to exhaustively process the frames in a camera until the object is found.
If the system examines a camera that does not contain the object, it has to process the entire camera feed before moving to the next camera.
% Both \dfs and \spatula use a static search plan that exhaustively looks at each camera feed until the object is found.
Both \dfs and \spatula do not take an \emph{adaptive query processing} approach for selecting the neighboring cameras.
So, the low camera prediction accuracy of these systems leads to the processing of several additional frames in each camera before moving to the next camera.
%
% \PC{Mention that we see a 99\% accuracy on AI city dataset
% with robust re-id models}

% Additionally, 
% To handle the task complexity, SOTA \reid models use multiple large deep neural networks in tandem~\cite{}
% to extract the features from the query object and the objects in the dataset.
% %
% So, each invocation of the \reid model is computationally expensive - on a server grade GPU,
% \reid model runs at a meager \PC{todo} fps.
% %
% On a large camera network, the number of \reid model invocations can be very large.
% %
% For example, consider a camera network with 30 cameras and 1000 frames per camera.
% %
% Assuming each frame contains 5 objects on average, the total number of \reid model invocations is 150,000.
% %
% So, the query processing time is \PC{todo} seconds - too slow for real-time applications.

% \squishitemize
% \item All pair comparison and DFS work but too many comparisons
% \item For example, 1000 cameras with 1000 frames each
% \item Parallel execution is an option but not always possible due to heavy \reid models that fully utilize the GPU.
% \squishend

\PPP{3. Lack of large-scale video datasets for re-id task}
The third challenge is the lack of large-scale multi-camera video datasets to evaluate \reid queries.
The largest existing real-world dataset for \reid~\cite{cityflow} suffers
from two limitations:
(1) The dataset is characterized by a small number of cameras (\textless20) in a confined geographical area, with each camera capturing less than 5 minutes of video footage.
(2) The dataset lacks the necessary camera history information essential for performing camera prediction tasks in systems like \spatula.
Collecting large-scale synchronized multi-camera video datasets for \reid poses challenges due to the need for access to real-world camera networks.
Further, the dataset must undergo rigorous checks to ensure that user privacy is preserved~\cite{dukeprivacy}.

% \squishitemize
% \item Another challenge of \reid is the lack of large-scale video data.
% \item Best option right now AI city - how many cameras and frames?
% \item Existing datasets have only a few cameras and frames.
% \item Challenging to collect data for \reid.
% \squishend

\PP{Our Approach}
In this paper, we present \sys, a VDBMS tailored for efficiently processing \reid queries.
\sys operates on the underlying camera graph using a novel adaptive query processing framework.
Specifically, \sys performs a probabilistic adaptive search over the graph network in small temporal windows using an exploration-exploitation strategy.
\sys utilizes the camera prediction probabilities at each camera to \textit{sample} the most probable neighboring camera at each time step while exploring a small temporal window of pre-defined duration.
Based on the outcomes of the current window, \sys dynamically updates the probabilities for the subsequent sampling iteration.
This way, \sys performs a dynamic incremental search over all the neighboring cameras to spot the object of interest efficiently.

Note that the sampling probabilities play a crucial role in determining the adaptive search process.
For example, a highly skewed probability array would result in a rigorous exploitation of the highest probable camera before exploring other cameras.
So, a highly accurate camera prediction model would result in often examining the correct cameras rigorously before examining any other cameras.
We observe that the most probable next camera at any given camera is dependent on the long-term trajectory of the object to reach this camera and not just the local camera trajectory that is used in \spatula.
The intuition behind this observation is that real-world object trajectories often contain long-term dependencies (\eg popular routes in a neighborhood).
So, \sys learns the long-term cross-camera correlations to predict the most probable next cameras.
Specifically, \sys trains a recurrent network on the historical trajectory information to learn the most probable next camera, given a sequence of cameras traversed so far.
Thus, the probabilistic search model combined with an accurate camera prediction model allows \sys to process \reid queries efficiently.
\cref{tab:baselines-overview} shows a qualitative comparison of \sys against existing systems for processing \reid queries.
%

%
% The adaptive search paradigm is inspired by the adaptive query processing literature in traditional database systems~\cite{aqp-survey}.
% %
% Specifically, the probabilistic data model can be envisioned as a routing policy within the eddies framework~\cite{eddies}.
% %

%
% At each camera, the n-gram models use the current object trajectory to predict the most likely neighboring cameras.
% %
% By processing the most likely cameras first, \sys efficiently avoids processing cameras unlikely to contain the query object.
% %
% We call this approach \textit{spatial filtering}.

% Once the optimal order of neighboring cameras is found, \sys uses temporal filtering to reduce the number of frames to be processed
% in each camera.
% %
% \sys uses statistical priors from the dataset to predict the most likely frames in a camera to contain the query object.
% %
% It also uses adaptive sampling to sample frames from the most likely frames.
% %
% The adaptive sampling strategy allows \sys to quickly arrive at the query object if it is present
% in the camera.
% %
% In this way, \sys iteratively applies spatial and temporal filtering to process \reid queries efficiently.

To evaluate \sys, we present a new large-scale synthetic benchmark for multi-camera \reid, based on the Carla simulator~\cite{carla}.
This benchmark allows spawning an arbitrary number of cameras in arbitrary locations and orientations in a game engine.
We develop a technique to simulate real-world traffic scenarios by mimicking real-world trajectory distributions.
The benchmark allows the simulation of different traffic patterns (\eg traffic density, distribution skew) to facilitate further research in VDBMSs.
%
%Users can also generate large-scale multi-camera video datasets on their custom camera network topologies automatically.
%

\begin{table}
    \begin{adjustbox}{width=\columnwidth,center}
    \begin{tabular}{cccccc}
    \toprule
                & \textbf{Multi Camera} & \textbf{Spatial} & \textbf{Temporal} & \textbf{Adaptive} &\textbf{Long-term}\\
                &  \textbf{Tracking} &   \textbf{Filtering}  & \textbf{Filtering} & \textbf{Processing} &  \textbf{Correlations} \\ \midrule
    Proxy Models~\cite{noscope, pp, blazeit} &  &  & \cmark & & \\
    MIRIS/OTIF~\cite{miris, otif} &  & & \cmark & & \cmark\\
    \dfs &  \cmark & \cmark & & & \\
    \spatula~\cite{spatula} &  \cmark & \cmark & \cmark & &\\
    \sys        & \cmark & \cmark & \cmark & \cmark & \cmark\\ \bottomrule
    \end{tabular}
    \end{adjustbox}
    \caption{
        \textbf{Systems for processing \reid queries.}
        \sys 
        (1) tracks object across cameras, 
        (2) performs spatio-temporal filtering, 
        (3) adaptively processes cameras in the network, and 
        (4) leverages long-term cross-camera correlations to accelerate the query.
        }
    \label{tab:baselines-overview}
\end{table}

%
% \notice{{\em Crown-jewel}: show-case our very best results, 
% easy to understand}.

\PP{Contributions}
In summary, the key contributions are:
%
% \notice{Use squishitemize and squishend to create compact itemized lists.
% squishenumerate and squishend for compact enumerated lists.}
%
\squishitemize
\item We highlight the limitations of SoTA spatio-temporal filtering systems in processing \reid queries over a large camera graph (\autoref{sec:introduction}).
We present a novel adaptive query processing system to overcome these limitations (\autoref{sec:system-overview}).
\item We model long-term cross-camera correlations using a recurrent network to improve the camera prediction accuracy (\autoref{sec:opt:camera-prediction}).
\item 
We develop a probabilistic adaptive search algorithm that processes camera feeds in small temporal windows and dynamically updates the probabilities using an exploration-exploitation strategy (\autoref{sec:opt:probabilistic-execution}).
\item We develop a novel \reid benchmark capable of generating multi-camera video datasets with diverse data distributions (\autoref{sec:eval-reid-benchmark}). 
\item
We evaluate \sys on synthetic benchmarks and city-scale real-world GPS trajectory datasets.
We show that \sys accelerates \reid queries by~4.7$\times$ on average compared to \dfs and by~3.9$\times$ on average compared to \spatula (\autoref{sec:evaluation}).
\squishend

\PP{Ethics statement}
%
%\rev{
%
% \PC{Add citations}
%
In this paper, we explore efficient multi-camera object re-identification,
which has both positive and negative applications.
On the positive side, this technology benefits urban planning, emergency protocols, sports analytics, and more.
However, there are potential risks, including privacy concerns with fine-grained traffic surveillance~\cite{ardabili2022understanding, almeida2022ethics}.
To mitigate these risks, we emphasize responsible and ethical use of the technology~\cite{board2019ai}.
We have taken concrete steps, including releasing an open-source synthetic benchmark for privacy-preserving dataset generation and using anonymized datasets in our analysis.
By emphasizing transparency and privacy protection, we strive to contribute to the responsible development of object re-identification technology.
%}
% \PP{Reproducibility} 
% %
% We have published the software artifacts at: \url{www.cc.gatech.edu}.
% \AJ{Need to update?}
%
%Do this only for double-blind conferences, where the reviewers may know our
%identity.

%% file: content/background.tex
\section{Background}
\label{sec:background}

\PP{Object re-identification}
Object re-identification (\reid) is the task of retrieving all the instances (target images)
of an object from a supplied query image.
When applied to multi-camera video feeds, \reid enables tracking objects over a network of cameras,
enabling multi-camera multi-target tracking (MCMT)~\cite{cityflow}.
% Traditionally, \reid augmented single-camera object tracking to enable multi-camera multi-target tracking (MCMT).
%
For example, a traffic monitoring expert may be interested in finding the route taken by a suspicious vehicle
in the city-scale road network.
\reid is challenging because of:
(1) significant variations in the scene, distance, and viewing angles of different cameras and
(2) occlusion due to other objects and camera placements.
%

% \squishitemize
% \item Object re-identification task
% \item Example query
% \item Different scenes/viewing angles/sizes
% \squishend

% \begin{figure}
%     \centering
%     \includegraphics[width=\columnwidth]{figures/background/reid-pipeline.pdf}
%        \caption{
%             %
%             Object Re-identification pipeline
%             %
%         }
%        \label{fig:reid-pipeline}
% \end{figure}

\begin{figure}[t!]
    \begin{adjustbox}{width=\columnwidth,center}
    \includegraphics[width=\columnwidth]{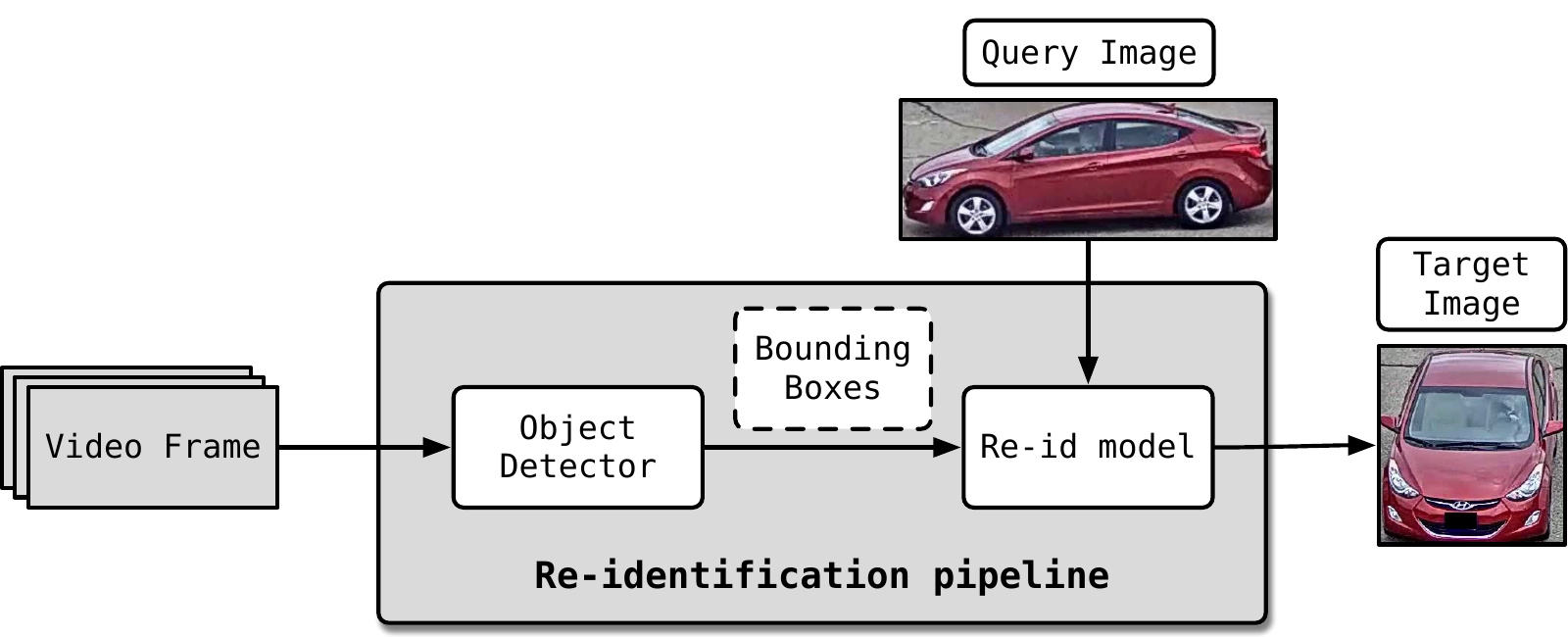}
    \end{adjustbox}
    \caption{A typical pipeline for object re-identification.}
    \label{fig:reid-pipeline}
\end{figure}

\PP{\reid pipeline} % \PC{Add more specific details of the models here - this text is already in introduction}
\cref{fig:reid-pipeline} shows the typical object re-identification pipeline
to retrieve the target image given a video frame and a query image.
An object detector first processes the video frame to find all the objects, and
returns the object bounding boxes.
The bounding boxes and the query image are processed by a \textit{\reid model}
to extract their respective feature vectors~\cite{reidbaseline}.
Subsequently, the similarity between each object's feature vector and the query image's
feature vector is computed using the standard Cosine or Euclidean distance
to identify the best matches.

% \squishitemize
% \item Object detection to get bounding boxes
% \item Use a \textit{re-id model} to extract features
% \item Feature vector similarity to match objects
% \squishend

\PP{\reid models}
The \textit{\reid model} determines the accuracy of the \reid pipeline above in finding correct matches.
%
% The model aims to generate the most expressive feature vector for the \reid task.
%
Two main classes of models have been used for \reid in literature.

\PPP{1. Feature engineering}
Early \reid models~\cite{zapletal2016vehicle} used hand-crafted techniques like histogram of oriented gradients (HOG)~\cite{hog} to generate feature vectors based on edge- and keypoint- detection.
However, these hand-crafted local features do not generalize to scene/view changes.
% - HOG, SiFT  % keep this very small and maybe not a separate PPP
%

\PPP{2. Deep neural networks} % this can be longer
Recent advancements in \reid involve deep neural networks (DNNs) to improve generalization.
%
% Recent works propose both part-based and global feature-based models for \reid.
%
Part-based models extract features for semantic partitions of the objects (\eg hood, trunk of a vehicle)
and aggregate these part-based features.
Global feature-based models, on the other hand, use global object features and
incorporate rigorous data augmentation to account for scene, orientation, and size changes.
The SoTA technique~\cite{reidbaseline} for detecting fine-grained object features uses a global feature-based model
with three large DNN backbones (variants of ResNet~\cite{resnet}) to generate a merged feature vector.
%
% These DNN backbones require many convolutional layers (\eg ResNet152~\cite{resnet}) for each feature vector.
%
So, feature extraction for \reid is computationally expensive, and up to 2$\times$
slower than standard object detection models such as YOLOv5~\cite{yolov5}.
%

% \squishitemize
% \item Feature extraction using DNNs.
% \item Data augmentation
% \item Computational complexity - object detection followed by feature extraction is expensive - give concrete numbers for both.
% \squishend
%

\PP{Challenges}
In the city-scale vehicle \reid task, time-synchronized videos are ingested from several cameras~\cite{cityflow}.
The objective is to find all the occurrences of the object in all the cameras given a query image.
This task has applications in various domains, such as traffic monitoring, urban planning, and sports analytics.
However, finding all the occurrences requires:
(1) detecting objects in each frame in all cameras,
(2) performing \reid feature extraction for all detected objects, and
(3) finding pairwise similarity (\eg cosine) between the source object and each detected object.
Given the vast number of detected objects, this task incurs significant computational costs.
For example, with 16 cameras capturing 5-minute sequences at 10 fps,
querying a single object requires 72,000 \reid pipeline invocations,
resulting in a total computation time of about 2 hours on a server-grade GPU.

\PP{\reid queries}
In many tasks, it is not necessary to find \textit{all the occurrences} of the object in each camera.
For instance, in time-sensitive applications like Amber alert tracking,
the primary objective is to \textit{spot the suspect vehicle} in the cameras (\ie we need one timestamp per camera
to track the vehicle in the camera network).
However, achieving a high recall rate is essential for the query to ensure exhaustive identification,
meaning that no cameras can be missed in the process.
We formulate this task as a \textit{\reid query}:
given a query image, the source camera, and the time-stamp of occurrence, find all the cameras
in the network that contain the query image.
%
% We consider such \reid queries as spatio-temporal queries.
% %
% The data is spatially (geographically) distributed and contains temporally consistent information.
%
To process such \reid queries efficiently, our goal is to identify the query object in all the cameras
that contain it using the fewest number of \reid pipeline invocations.
%

% Specifically, our goal is to track the object in the camera network while efficiently
% handling camera failures, algorithmic errors, and model errors.
% \squishitemize
% \item Queries over spatiotemporal video data
% \item Define "occurrences" - Spotting an object in a camera feed and moving on to next
% \item Naive all-pair retrieval for multi-camera tracking
% \item Cost of processing re-id queries in terms of pairwise comparisons - specify a one-to-one mapping between the
% number of pairwise comparisons and the actual cost - use a table with cost?.
% \item Goal of \sys is to reduce the number of pairwise comparisons.
% \squishend
%
% %

% \subsection{Action Localization}
% \label{sec:overview-action-localization}
% %
% Action queries focus on  (1) action classification which answers  \textit{what}
% actions are present in a long, untrimmed video, and (2) temporal boundary
% localization for \textit{where} these actions are present within the video.
% %
% We refer to the combined task of detecting, and localizing actions in videos as
% action localization (AL).

%

%% file: content/solution.tex
%\section{Executing \reid Queries}
\section{System Overview}
\label{sec:system-overview}

%\PP{Problem Setup}
%
\sys assumes that a camera network topology is given as an unweighted graph $G=(V,E)$.
In this graph, each vertex $v \in V$ represents a camera and each edge $e \in E$ represents
a connection between two adjacent cameras.
Note that adjacent cameras in the graph might be physically far away from each other (\eg highway exits).
We assume that each camera captures videos of length $T$ frames
synchronously at a fixed frame rate.
We assume that the video feeds from all the cameras are recorded and sent to a central server.
These video feeds are updated periodically.

The input to \sys is a query object bounding box $q$, a source camera $c_q \in V$, and the time stamp $t_q \in [1, T]$ when the query object was found in $c_q$.
%
% The query image $q$ is available at runtime.
%
% Additional query images can also be provided to \sys.
%
\sys's goal is to find the occurrences of the query object in the camera network rapidly.
%
% \rev{
% %
% Recall that we move to the next camera as soon as we spot the object once in a given camera.
% }
%
Specifically, \sys returns the target object bounding box $q'$ along with the associated camera and timestamp pair $<c_q', t_q'>$, one for each camera that contains the query object.
% a set of candidate cameras $V' \subseteq V$ and the time stamp $t_v \in [1, T]$ at
% which the object was identified for each candidate camera $v \in V'$.
%
Note that we are only interested in spotting the vehicle in each camera feed.
So, we only need to find one frame in each camera that contains the object.

\begin{figure}
    \centering
    \includegraphics[width=\linewidth]{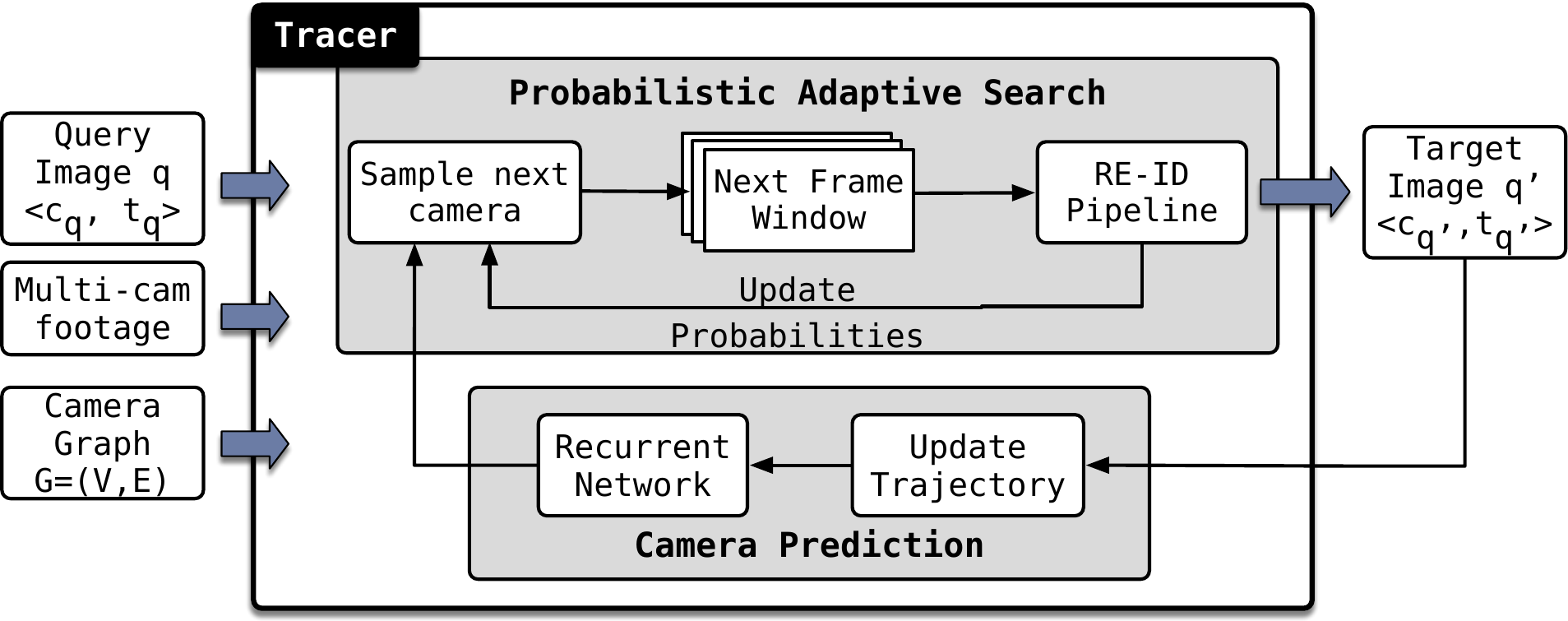}
       \caption{
            System architecture of \sys.
        }
       \label{fig:system-architecture}
\end{figure}

% \PC{Add references to section 4}
%
\cref{fig:system-architecture} shows the system architecture of \sys.
As discussed in the problem setup, \sys takes a query image $q$ with the source camera and time stamp $<c_q, t_q>$ as input.
It operates on a synchronous multi-camera video feed of length $T$ frames collected from the camera graph $G$.
\sys consists of two novel components that contribute to its high performance.

\PP{Probabilistic Adaptive Search}
%
%\rev{
%
\sys employs a probabilistic adaptive search strategy to achieve high recall efficiently.
As shown in~\autoref{fig:system-architecture}, \sys dynamically selects the next camera to process at each time step from candidate cameras within the graph.
The selection involves sampling from a probability distribution generated by the camera prediction module (\autoref{sec:opt:camera-prediction}).
Once a camera is chosen, \sys incrementally processes a window of its frames, executing the \reid pipeline~(\cref{fig:reid-pipeline}).
If the query object is  detected, \sys returns the target image $q'$, camera $c_q'$, and timestamp $t_q'$.
Additionally, \sys also appends $c_q'$ to the current trajectory and uses it as input to the camera prediction module.
However, if the query object is not found, \sys updates the probabilities using an exploration-exploitation rule (\autoref{sec:opt:probabilistic-execution}) and proceeds to the next sampling round.
This allows \sys to retrieve all relevant cameras by only incrementally exploring video feeds.
%}
%

\PP{Camera Prediction}
%
%\rev{
%
\sys features a camera prediction module that accurately predicts the next camera to examine at each time step (\autoref{sec:opt:camera-prediction}).
The module's primary output is the probability of finding the object in each of the neighboring cameras.
On dense camera networks with a high degree, the choices for the next camera are large.
In such cases, we find that local cross-camera correlations (\autoref{fig:intro-spatula}) are insufficient for accurate camera prediction.
To address this issue, the camera prediction module in \sys trains a recurrent neural network (\autoref{sec:opt:recurrent-network}).
The trained network takes the trajectory traversed thus far as input and outputs probabilities for finding the object in each neighboring camera.
By utilizing long-term trajectories, \sys effectively captures cross-camera correlations even between distant cameras within the camera graph, enhancing the accuracy of camera prediction.
%}
% \rev{
% Combining both the modules, the query cycle in \sys is as follows:
% %
% At each camera, the camera prediction module takes the current trajectory as input and outputs probabilities.
% %
% The adaptive search module then performs multiple sampling rounds during which, different neighboring cameras are searched in incremental windows.
% %
% If the target object is found, the search moves to the next camera and its neighboring.
% %
% If the target object is not found in any of the neighbors, the search terminates and the current trajectory is the final output trajectory.
% }

%\rev{
%
Both the camera prediction and the probabilistic adaptive search modules operate in conjunction to enhance the performance of \sys.
Specifically, the notable enhancement in accuracy provided by the camera prediction module (\ref{sec:eval-camera-prediction-model}) allows \sys to frequently select the correct camera with confidence.
In such scenarios, the search module facilitates effective exploitation of the identified camera to quickly identify the target object without unnecessary delays caused by incorrect cameras.
In the event that the camera prediction module produces inaccurate results, the search module promptly adjusts to a random exploration strategy to prevent being trapped by incorrect cameras.
\sys thus efficiently executes the \reid query over the camera network using a combination of probabilistic adaptive search and accurate camera prediction.

\section{Problem Formulation}
\label{sec:opt:problem-formulation}
%
% \PC{Revisit and update}
% %
% In this section, we describe the optimizations employed in \sys to accelerate \reid queries.
% %
% We begin by formulating the search space reduction problem in~\autoref{sec:opt:problem-formulation}.
% %
% Next, we describe the camera prediction problem within the context of executing \reid queries and
% discuss the limitations of na\"ive solutions in \autoref{sec:opt:camera-prediction}.
% %
% To address these limitations, we propose capturing long-term cross-camera correlations in \autoref{sec:opt:long-term-correlations} and demonstrate the inadequacy of statistical models in accurately capturing these correlations.
% %
% To overcome this, we introduce a recurrent network in \autoref{sec:opt:recurrent-network} to model long-term correlations accurately.
% %
% Lastly, we discuss the probabilistic adaptive search plan in \autoref{sec:opt:probabilistic-execution}.
% %

% \subsection{Problem Formulation}
% \label{sec:opt:problem-formulation}
%
We begin by formally defining the problem statement.
As discussed in~\autoref{sec:system-overview}, we are given a camera network $G = (V, E)$
and a query image q with source camera, timestamp pair $<c_q, t_q>$.
%\
The na\"ive approach of brute-force searching through all the cameras in the network
becomes infeasible due to the large number of cameras and the substantial number
of frames within each camera.
The goal is to output all the cameras that contain the query object in the
camera network $G$ while examining as few frames as possible.
The search space reduction problem can be formally defined as follows:
\begin{problem}[Search Space Reduction]
\label{prob:search-space-reduction}
Given a camera network $G = (V, E)$, a query object bounding box $q$, along with a source camera and timestamp pair $<c_q, t_q>$, the objective is to
\textit{minimize the number of frames examined} to retrieve all the cameras $V' \subseteq V$ that contain the query object.
\end{problem}

Leveraging our knowledge of the underlying graph structure, we can limit the search space to the
reachable cameras from $c_q$ and employ a graph traversal technique (\dfs) to locate the target object.
\dfs traverses the camera network $G$ from the source camera $c_q$ and iteratively explores the neighboring
cameras until it locates the target object (\cref{fig:intro-graph-search}).
However, this approach requires examining all the frames within each camera before moving to the next camera.
Additionally, it requires examining several additional cameras since it explores the neighboring cameras in random order.
%
% Additionally, it requires examining all the frames within each camera.
%
The oracle solution to this problem (\oracle), which has access to the ground truth object cameras and timestamps,
can retrieve the subset of cameras $V'$ by examining $|V'|$ frames (one frame per camera).
The \oracle can be visualized as a traversal algorithm that always selects the correct neighboring camera to explore
and always finds the object in the first frame that it examines.
This is the best possible solution to the search space reduction problem and serves as the upper bound for
any solution.
A close-to-optimal solution to ~\cref{prob:search-space-reduction} should examine only a small subset of frames within each camera and predict
the neighboring camera order with high accuracy.
%

% \PC{Limitations:
% (1) we assume that there are no errors such as missing detections
% (2) we can handle loops due to incremental search but not back and forth (U-turns)}

\PP{Problem Scope}
%
%\rev{
%
Objects often appear in multiple consecutive frames in a camera due to temporal continuity.
However, recall that our primary objective is only to spot the object in each camera.
Once we detect the object in a camera, we output the target bounding box $q'$, along with the corresponding camera and timestamp tuple $<c_q', t_q'>$, and proceed to search for the object in the next camera.
%
%}

%\rev{
%
Additionally, objects may appear multiple times in the same camera.
Our system natively supports detecting objects reappearing in the same camera after several camera hops (\eg vehicle loops).
However, rapid oscillations of objects between two cameras lead to poor recurrent model convergence.
Sparse camera networks at city scale rarely exhibit such trajectory patterns~\cite{cityflow}.
So, \sys currently does not support it.
% Our analysis of GPS trajectory datasets (\autoref{sec:eval-setup}) indicates that such trajectories are primarily caused by noisy GPS tracking.
%
%
%}

\begin{figure*}[t!]
    \centering
    \begin{subfigure}[b]{0.2\linewidth}
        \centering
        \includegraphics[width=\linewidth]{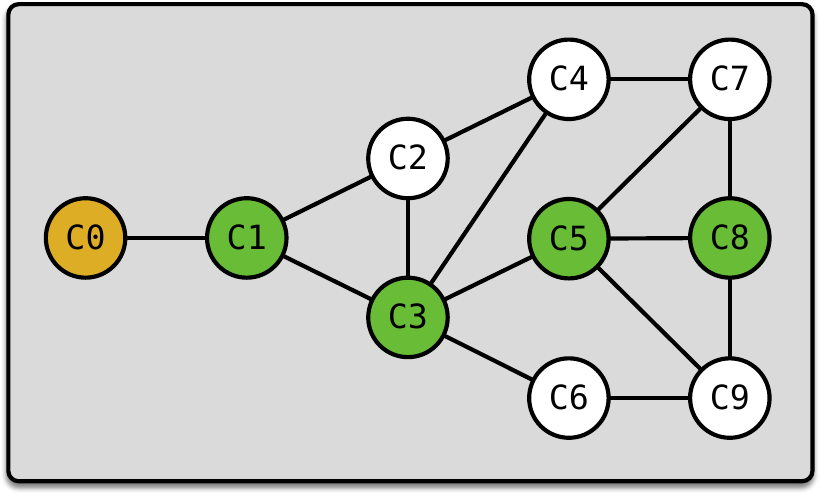}
           \caption{
                \textbf{Illustrative camera network topology --} A query image found in camera C0 at time step t=0
                is used to track the object across the network.
            }
           \label{fig:illustrative-example-topology}
    \end{subfigure}
    \hfill
    \begin{subfigure}[b]{.79\linewidth}
        \tiny 
        \centering
        \resizebox{0.9\textwidth}{!}{%
        \begin{tabular}{ccccc} % @{\hspace{3\tabcolsep}} c @{\hspace{3\tabcolsep}}c @{\hspace{3\tabcolsep}} c }
         &
        \includegraphics[width=0.05\linewidth]{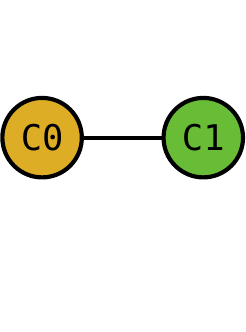} &
        \includegraphics[width=0.05\linewidth]{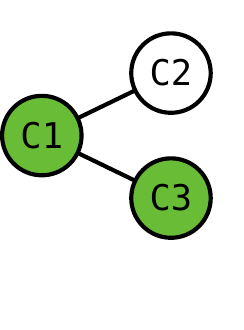} &
        \includegraphics[width=0.05\linewidth]{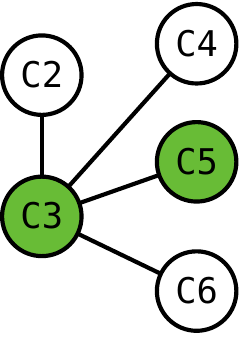} &
        \includegraphics[width=0.05\linewidth]{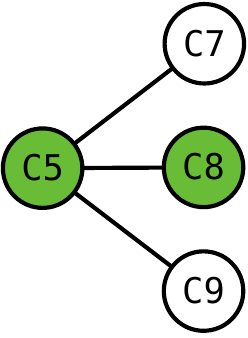}
        \\ \hline
        \tiny \textbf{Time Step} & t=1 & t=2 & t=3 & t=4 \\ \hline
        % $\#$ Neighbors & 1 & 2 & 4 & 5 \\
        \dfs search order & C1 & C2, C3* & C4, C2, C5*, C6 & C9, C7, C8* \\
        \spatula input & C0 & C1 & C3 & C5 \\
        \spatula search order & C1 & C3*, C2 & C6, C2, C4, C5* & C9, C8*, C7 \\
        \sys input & [C0] & [C0, C1] & [C0, C1, C3] & [C0, C1, C3, C5] \\ 
        \sys search order & C1 & C3*, C2 & C6, C5*, C2, C4 & C8*, C7, C9 \\ \hline
        % Prev. timestamp & f=10 & f=125 & f=410 & f=600 \\
        % Search window & (100, 250) & (350, 450) & (500, 800) & (900, 1000) \\
        % Ground truth window & (150, 160) & (345, 355) & (700, 705) & (970, 990) \\
        % Found frame & f=155 & f=410 & f=600 & f=920 \\ \hline
        % $\#$ Frames processed & 6 & 8 & 5 & 4 \\ 
        % Frame Number & f=1 & f=65 & f=97 & f=101 \\
        % Frame Number & f=1 & f=65 & f=97 & f=101 \\
        % Frame Number & f=1 & f=65 & f=97 & f=101 \\
        \textbf{Total cameras processed} &  &  &  &  \\
        \dfs/\spatula/\sys & 1/1/\textbf{1} & 2/1/\textbf{1} & 3/4/\textbf{2} & 3/2/\textbf{1} \\ \hline
        \end{tabular}
        }
        \caption{
            \textbf{Camera prediction in \sys on the given graph --}
            \sys efficiently processes \reid queries by accurately predicting the order of processing the neighboring cameras.
        }
       \label{fig:illustrative-example-table}
    \end{subfigure}
    \caption{ 
    	An illustrative example of \sys's camera prediction module
     }
     \label{fig:illustrative-example}
\end{figure*}

\section{Camera Prediction}
\label{sec:opt:camera-prediction}
We first describe the camera prediction module of \sys.
%
% To accurately predict the neighboring camera order, we draw inspiration from \spatula~\cite{spatula}
% and leverage the spatial-temporal correlations between cameras.
% %
% Although we make use of temporal correlations, the main emphasis of
% this paper is on spatial correlations.
% %
% This choice is motivated by the high recall setting discussed in \autoref{sec:introduction}.
%
%\PC{Should we add this temporal filtering disclaimer earlier in the paper itself?} -- No
%
Let us define the event of finding the query object $q$ in the camera $v$ as $v_{q}$.
% %
At each camera $u$, our goal is to estimate the probability of finding the query object $q$ in the set of
neighboring cameras $ v \in V_{n}$, written as $P(v_{q})$.
% %
To maintain brevity, we will omit the object $q$ from the notation throughout the remainder of this paper
and utilize $P(v)$ in place of $P(v_{q})$.

\subsection{Local correlations}
\label{sec:opt:local-correlations}
To estimate this probability, \spatula computes the Maximum Likelihood Estimate (MLE) of the probability
using the historical frequency count of objects that traveled from $u$ to each of the neighboring cameras
$v \in V_n$.
\[ P_{MLE}(v) = \frac{C(v)}{N} \]
where $C(v)$ is the frequency count of camera $v$ in the historical object tracks and \( N = \sum_{u \in V_n}{C(u)} \)
is the total count of all the objects in all the neighboring cameras.
We can compute $P_{MLE}(v)$ for all the neighboring cameras and generate a probability array over the
neighboring cameras.
Thus, this approach generates a search order for the neighboring cameras by utilizing the localized historical information
to identify the neighboring cameras that have been visited most frequently.
However, the drawback of this approach is that a simplistic frequency estimate that only looks at the localized camera
information lacks the necessary granularity to estimate the probability accurately.
Our experiments show that this approach achieves less than 50\% accuracy in predicting the neighboring camera order
even on sparse camera networks (\autoref{sec:eval-camera-prediction-model}).

\subsection{Long-term Correlations}
\label{sec:opt:long-term-correlations}
To improve the accuracy of camera prediction, we propose a novel approach that utilizes the long-term correlations
between cameras to accurately estimate the probability of finding the query object in the neighboring cameras.
For example, consider a scenario where two neighboring cameras, A and B, have been visited an equal number of times
in the past.
In this case, relying solely on the frequency estimate would treat both cameras as equally probable,
even though the historical object trajectories may indicate that the object is more likely to be found in camera A.
Our intuition for this approach is that object trajectories in the real world are likely to follow
certain path distributions.
For example, in a city, there are likely to be specific traffic routes that are more popular
than others due to ease of access.
Furthermore, cities typically comprise hotspots for both the source and destination points,
which leads to certain trajectories being more probable than others.
For example, an analysis of the New York City taxi dataset~\cite{nyc-taxi} reveals that
80\% of the trips originate or terminate at 10\% of the total locations (\autoref{fig:traffic-distribution}).
More formally, let us denote the path taken by the object to reach the current camera $u_{k-1}$ as \( \{u_{1}, u_{2}, \ldots, u_{k-1}\} \),
where $k-1$ is the path length so far from the source vertex $u_1$.
Then, our objective is to estimate the probability of finding the object $O$ in neighboring camera $u_k$ of $u_{k-1}$ conditioned on the path taken so far,
written as \( P(u_k|u_{1}, u_{2}, \ldots, u_{k-1}) \).

\subsection{N-gram approach}
\label{sec:opt-n-gram}
An obvious choice to estimate this conditional probability is the n-gram model~\cite{ngram}.
N-gram models are a popular technique used in natural language processing to estimate the probability of
a word given the previous n-1 words.
Consider the object trajectory in the camera graph as a sequence $\overline{u}$ of $k$ cameras from a starting camera $u_1$.
%
% Directly estimating the probability of the entire sequence from historical object trajectories becomes infeasible due to the
% exponential growth in the number of sequences, resulting in a significant portion of sequences being absent from the dataset.
% %
% To improve the estimation of this probability, we employ the chain rule to express the probability as follows:
% %
% \begin{align*}
% \begin{aligned}
%  P(\overline{u}) = {} & P(u_k | u_{k-1}, u_{k-2}, \ldots, u_1) \\
%                             & \times P(u_{k-1} | u_{k-2}, u_{k-3}, \ldots, u_1) \\
%                             & \times \ldots \\
%                             & \times P(u_2 | u_1)
% \end{aligned}
% \end{align*}
%
Using the Markov assumption, n-gram models simplify the probability estimation to use a fixed number of previous cameras to predict the next camera.
For a given n, the probability is defined as:
\begin{equation}
\label{eqn:n-gram}
    \begin{aligned}
     P(\overline{u}) = {} & P(u_k | u_{k-1}, u_{k-2}, \ldots, u_{k-n+1}) \\
                                & \times P(u_{k-1} | u_{k-2}, u_{k-3}, \ldots, u_{k-n}) \\
                                & \times \ldots \\
                                & \times P(u_2 | u_1)
    \end{aligned}
\end{equation}
%
% \PP{Training the n-gram model}
%
For n=1 (\textit{unigram}), this probability presumes the strong independence condition and reduces to the simple frequency estimate used in \spatula (\autoref{sec:opt:local-correlations}):
\[ P(\overline{u}) = \prod_{i=1}^{k}{P(u_i)} \]
To train general n-gram models, we must compute the probability \( P' = P(u_k | u_{k-1}, u_{k-2}, \ldots, u_{k-n+1}) \) in~\cref{eqn:n-gram} using all the available sequences of length $n$ in the dataset.
More specifically, for each neighbor $u_k$ of camera $u_{k-1}$, we need to find the MLE of $P'$.
For each neighbor $u_k$, this probability can be estimated as:
\begin{align*}
    \begin{aligned}
    P'_{MLE}(u_k) = {} & \frac{P(u_{k-n+1}, \ldots, u_{k-1}, u_k)}{P(u_{k-n+1}, \ldots, u_{k-1})} \\
                                            = & \frac{Count(u_{k-n+1}, \ldots, u_{k-1}, u_k)}{Count(u_{k-n+1}, \ldots, u_{k-1})}
    \end{aligned}
\end{align*}
In other words, for each neighboring camera, we must compute the frequency of occurrence of the sequence $<u_{k-n+1}, \ldots, u_{k-1}, u_k >$ and the sequence $<u_{k-n+1}, \ldots, u_{k-1}>$.
To compute these quantities for any given sequence, we use the historical data of the vehicle trajectories.
%
% Consider a historical record of vehicle trajectories in the dataset.
% , each of the form $ X_t = \{O_{u_1}, O_{u_2}, \ldots, O_{u_{t}}\} $,
% where $t \rightarrow \{1,\ldots,T\} $.
% %
We traverse each trajectory in a sliding window fashion to update the counts for each $n$ length sequence.
The first $n-1$ cameras in the sequence are used as the index to update the frequency count for the $n^{th}$ camera.
At inference time, we use the last $n$ cameras in the current trajectory of the object to index into the n-gram model and generate the list of probabilities for the neighboring cameras.
% We show the algorithm in~\cref{alg:n-gram-training}.
%
% \input{algorithms/fixed-n-gram.tex}
%
% \PC{Go over the algorithm line-by-line}
% \squishitemize
% \item Given a trajectory $X_t$, the n-gram model can be updated in a sliding window fashion over window of length n.
% \item Do it in a loop over all the trajectories. \PC{check how to use likelihood estimate instead of just counting}
% \item Discuss the trade-off between n and the model accuracy as a segway to the variable n-gram model.
% \squishend
% %

\subsection{Recurrent Network}
\label{sec:opt:recurrent-network}
The use of n-gram models to capture long-term spatial correlations is suboptimal due
to the lack of prior knowledge on the optimal length of n.
Moreover, the optimal length of $n$ might vary across different segments of the trajectory.
For example, consider a vehicle trajectory that starts at a highway and ends at a city.
The optimal length of $n$ may vary between the highway and city segments of the trajectory,
as the traffic distribution in these two segments is different.
Consequently, our experiments show that the n-gram model improves the prediction accuracy
by only 8\% on average over the frequency estimate used in \spatula (\autoref{sec:eval-camera-prediction-model}).

To address this issue, we propose the use of recurrent neural networks (\rnn) to
capture the long-term spatial correlations.
\rnn{s} are a class of neural networks that are tailored for sequential data.
They use feedback connections to pass information from one time step to the next~\cite{rnn-seminal}.
This allows them to capture long-term dependencies in the data.
\rnn{s} preserve information across time steps by maintaining a hidden state $h_t$ that
serves as a proxy representation of the past input sequences.
The ability of RNNs to handle variable-length sequences makes them highly suitable for our problem.
The \rnn model improves the prediction accuracy by 25\% on average
over \spatula (\autoref{sec:eval-camera-prediction-model}).

\PP{Illustrative example}
\cref{fig:illustrative-example} shows an example of \sys's camera prediction pipeline.
Consider the 10 camera topology in \autoref{fig:illustrative-example-topology}.
Given a query image (\eg vehicle bounding box) captured in camera C0
at time $t=0$, \cref{fig:illustrative-example-table} shows the processing states of \dfs, \spatula, and \sys at each time step.
At $t=1$, all systems pick C1 as the next camera, as it is C0's only neighbor.
At $t=2$, C2 and C3 are C1's neighbors, with the object in C3.
\dfs randomly predicts the search order C2, C3.
\spatula uses the current camera C1 to predict the search order C3, C2.
\sys uses the current trajectory [C0, C1] as input to the \rnn and predicts
the search order C3, C2.
Since the object is in C3, \dfs processes one additional camera, while \spatula and \sys process no additional cameras.

At $t=3$, \dfs, \spatula, and \sys explore 2, 3, and 1 additional cameras, respectively.
Both \dfs and \spatula are inaccurate in this step - \dfs due to the random prediction order, and \spatula because it only considers the localized camera history at C3.
In contrast, \sys leverages the full trajectory [C0, C1, C3] to predict a more accurate order.
Note that \sys also incurs additional cost here, albeit lower than the other baselines.

Finally, at $t=4$, \dfs and \spatula explore 2 and 1 additional cameras, respectively.
\sys's \rnn correctly predicts the next camera using the current trajectory [C0, C1, C3, C5].
Despite mispredicting the camera in the previous time step, the \rnn's current input remains the ground-truth trajectory due to the 100\% recall constraint.
%
%This allows \sys to achieve high accuracy as the trajectories used for inference follow the same distribution as the ground truth trajectories used for training.
%
After 4 time steps, \sys selectively processes 5 of 10 cameras, while \dfs and \spatula process 9 and 8 cameras.

\PP{Training the \rnn model}
We use an LSTM network~\cite{lstm} with one hidden layer (128 units)
as the \rnn model.
For training, a batch of sequences are fed as the input,
while the corresponding sequences, right-shifted by 1, are used as labels.
The RNN is thus trained to predict the next camera in the sequence effectively,
regardless of the sequence length.
A fully-connected layer on the final hidden state produces the
neighboring camera probability distribution.
The model is trained using the Adam optimizer~\cite{adam} with a learning rate 0.001.
During inference, the trained RNN model iteratively predicts the next camera
by incrementally extending the input trajectory length,
as illustrated in~\autoref{fig:illustrative-example-table}.
%
% As we operate in the 100\% recall setting, we consistently utilize the actual sequence of
% cameras traversed by the vehicle as the input to the RNN rather than relying on the predicted sequence.

\PP{Training data acquisition}
For real-world deployment, the ground truth trajectories for training can be generated offline in a time-insensitive manner by running the \reid pipeline on historical video feeds.
Additionally, the inference trajectories from recent queries can be used as ground-truth training trajectories for subsequent queries.
This allows \sys to periodically update the prediction model and readily handle traffic distribution drift~\cite{odin}.
We plan to explore more efficient ways of handling distribution drift in future work.
% This data remains useful for several thousand queries before requiring model
% retraining.
%

\section{Probabilistic Adaptive Search}
\label{sec:opt:probabilistic-execution}
With the \rnn model, we can predict the next camera in the sequence with high accuracy.
However, the RNN model does not achieve 100\% accuracy, and it might mispredict the order of the camera search.
For example, consider the scenario in~\cref{fig:illustrative-example-table}.
The \rnn model mispredicts the camera search order at time step 3,
and starts the search at camera C6 instead of C5.
An exhaustive search model searches all the frames in a predicted
camera, and so this misprediction would result in processing all the frames in C6 before
searching C5.

To address this concern, we present a probabilistic adaptive search model that analyzes
camera frames in small temporal windows, guided by the probability distribution predicted
by the \rnn model.
More precisely, the next camera is sampled from the neighboring cameras using the probability
distribution predicted by the \rnn model.
Subsequently, we search within a fixed window of frames in the sampled camera.
The window size is tuned once per camera network based on the average duration of the objects within
a camera view.
In case the object is not detected in the sampled camera, we update the probability
distribution using an exploration-exploitation strategy that balances the exploration of
new cameras with the exploitation of the predicted camera.

\PP{Probability Update Algorithm}
Let the initial probability array be $P = [p_1, p_2, \ldots, p_n]$,
where $p_i$ is the probability for the $i^{th}$ neighboring camera.
After exploring a camera $i$ in a sampling round, the updated probability array
$P' = [p'_1, p'_2, \ldots, p'_n]$ can be calculated as follows:
$$ p'_i = \alpha \cdot p_i $$
$$ p'_j = p_j + \frac{{p_i*(1 - \alpha)}}{{n - 1}} \quad \text{for } j \neq i $$
In this update algorithm, $p'_c$ represents the updated probability for each neighboring
camera c, $\alpha$ is the exploration factor, and $n$ is the total number of
neighboring cameras.
This update algorithm can be interpreted as a way to balance the exploration of new cameras
with the exploitation of the predicted most probable camera.
The exploration factor $\alpha$ serves as a hyperparameter that can be tuned to achieve the
desired balance between exploration and exploitation.
A higher value of $\alpha$ results in a slower reduction of the probability of the
most probable camera, and thus, a higher exploitation rate.

The intuition behind this update algorithm is that a highly accurate
\rnn model often produces a skewed probability distribution.
When the underlying learned model is highly accurate, the value of $\alpha$
is set close to 1 to preserve the original probability distribution to a certain degree.
When the object is present in the most probable camera, the rigorous exploitation strategy
enables \sys to locate the object within the initial sampling rounds, during which the
probability of the most probable camera remains high.
However, if the object is not found in the most probable camera, the exponential reduction
in the probability of the most probable camera leads to a more balanced probability
distribution within a few subsequent sampling rounds.
In cases of less accurate models (\eg \spatula), 
$\alpha$ can be set to a lower value to promote increased exploration of neighboring cameras.

\begin{figure}[t!]
    \huge
    \centering
    \resizebox{\columnwidth}{!}{%
    % \addtolength{\tabcolsep}{-1pt}
    \begin{tabular}{ccccc} % @{\hspace{3\tabcolsep}} c @{\hspace{3\tabcolsep}}c @{\hspace{3\tabcolsep}} c }
    % \huge
     &
    \includegraphics[width=0.3\linewidth]{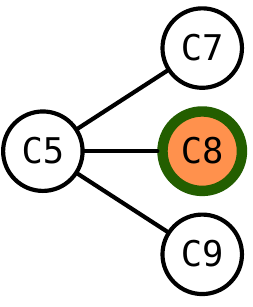} &
    \includegraphics[width=0.3\linewidth]{figures/overview/illustrative-example/t1-updated.pdf} &
    \includegraphics[width=0.3\linewidth]{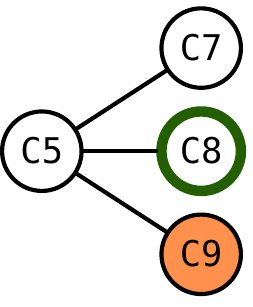} &
    \includegraphics[width=0.3\linewidth]{figures/overview/illustrative-example/t1-updated.pdf}
    \\ \hline
    \textbf{Sampling round} & 1 & 2 & 3 & 4 \\ \hline
    Probabilities & [0.1, 0.8, 0.1] & [0.2, 0.6, 0.2] & [0.3, 0.4, 0.3] & [0.4, 0.5, 0.1] \\
    Sampled camera & C8 & C8 & C9 & C8 \\
    Search window & (1, 75) & (76, 150) & (1, 75) & (151, 225) \\ \hline
    \textbf{Total Frames} & 75 & 150 & 225 & 255 \\ \hline
    \end{tabular}
    % \addtolength{\tabcolsep}{1pt}
    }
    \caption{
        \textbf{Probabilistic execution model in \sys --}
        \sys dynamically updates the sampling probabilities for each neighboring camera and adaptively
        expands the search in each neighboring camera feed to process \reid queries efficiently.
        The object is eventually located in camera C8 during the frame interval (180, 220).
    }
   \label{fig:illustration-probabilistic-execution}
\end{figure}

\PP{Illustrative example}
\cref{fig:illustration-probabilistic-execution} shows an illustrative example of the probabilistic adaptive search model of \sys.
Specifically, we focus on the time step $t=4$ in ~\cref{fig:illustrative-example-table}, where the current camera is C5 and the object is located in C8.
To enhance clarity of presentation, assume that the search process is initiated from frame 1 and the object is located in the frame interval (180, 220) in C8.
The search window is fixed at 75 frames, and the probabilities are updated after each sampling round using the update algorithm discussed above.
At $t=4$ in~\cref{fig:illustrative-example-table}, the input to the recurrent network is the trajectory [C0, C1, C3, C5].
Utilizing this trajectory, the recurrent network produces the probability array [0.1, 0.8, 0.1] for the neighboring cameras [C7, C8, C9], respectively.
During the initial sampling round, the system selects camera C8 and searches within the window (1, 75).
As the object is not detected in C8, the system updates the probabilities to [0.2, 0.6, 0.2].
In the second round, the system again samples camera C8 and performs a search within the window (76, 150) but fails to locate the object.
The probabilities are now updated to [0.3, 0.4, 0.3].

In the third round, the system samples camera C9 since the probabilities are more uniform, and initiates the search in the window (1, 75).
It is notable that the dynamic probability update algorithm enhances the chances of exploring neighboring cameras when the system fails to locate the object in the most probable ones.
As the object is not detected, the probabilities are updated to [0.4, 0.5, 0.1], and the system selects camera C8 during the fourth round.
Subsequently, the system identifies the object in frame 180 and outputs the target object's location and timestamp before transitioning to the next camera.

%% file: content/evaluation.tex
\section{Multi-camera \reid Benchmark}
\label{sec:eval-reid-benchmark}
An essential challenge for optimizing and evaluating \reid queries is the lack of a large-scale multi-camera video dataset.

\PP{Limitations of existing datasets}
The largest real-world multi-camera dataset is CityFlow from the Nvidia AI City Challenge~\cite{cityflow}.
CityFlow includes 3 hours of traffic videos from 40 cameras at 10 intersections in a US city.
The videos are 1-5 minutes long, temporally unsynchronized, and hence unsuitable for temporal analysis.
The most extensive map has under 10 intersections and 337 unique vehicles, severely limiting its usefulness to the \reid task.
A key challenge in obtaining large real-world multi-camera \reid datasets is privacy concerns.
Real-world datasets must be curated to remove personally identifiable information, such as license plates, before public release.
For example, the Duke-MTMC dataset was released and then retracted due to privacy issues~\cite{dukeprivacy}.

\PP{Synthetic Datasets}
To  circumvent these limitations of real-world datasets, researchers have recently proposed synthetic datasets such as VisualRoad~\cite{visualroad} and Synthehicle~\cite{synthehicle}.
However, they do not meet the following requirements for evaluating \reid queries:

\PPP{1. Long synchronized multi-camera video feeds}
The dataset should comprise video feeds from multiple synchronized cameras, ensuring that vehicles adhere to temporal constraints, and the video feeds should be sufficiently long to capture multiple vehicle tracks across the network.
VisualRoad is not suitable for the \reid task due to its inability to generate temporally synchronized multi-camera videos.

\PPP{2. Realistic data}
The simulated data must closely follow the real-world data distribution and behavior.
Specifically, the dataset should contain (1) realistic vehicle movements, such as adherence to traffic rules (2) realistic traffic patterns, including traffic hotspots and long-tail distributions.
However, VisualRoad and Synthehicle are not capable of generating such datasets.
% Specifically, the dataset should provide sufficient variation in terms of vehic
% However, this is difficult to acquire in the real world due to the large cost of deploying and collecting the videos.
% %
% Additionally, real-world videos have significant privacy concerns.
%

% \squishitemize
% \item Multiple cameras and synchronized video feed.
% \item Map customization and control over video sizes.
% \item Realistic traffic patterns - reference to~\cref{fig:eval-sample-images}.
% \squishend
\begin{figure}
    \centering
    \begin{subfigure}[b]{0.49\columnwidth}
        \centering
        \includegraphics[width=\linewidth]{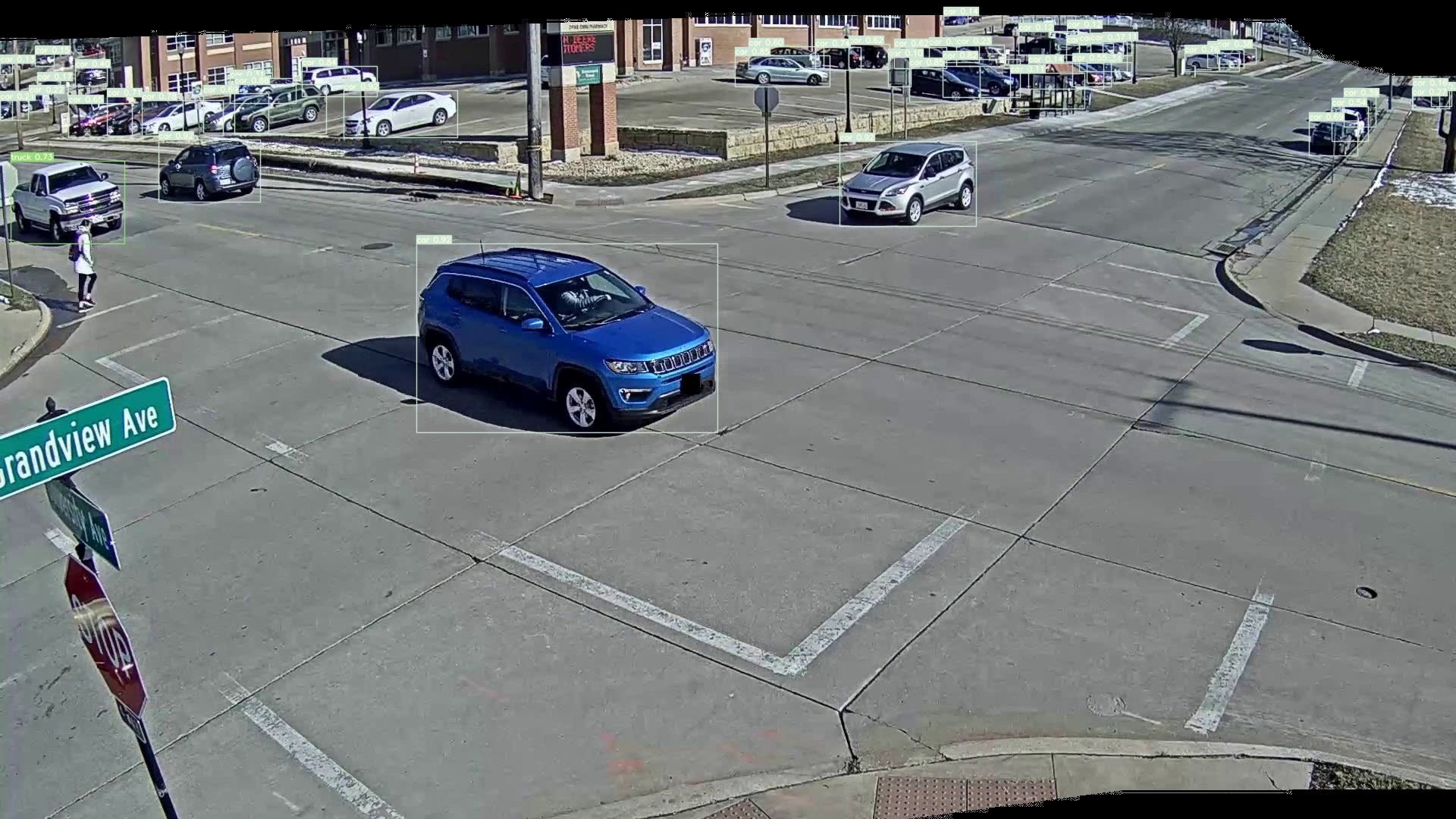}
           \caption{
                 Sample frame from the AI city dataset - The cameras are typically deployed at intersections.
            }
           \label{fig:eval-sample-images-ai-city}
    \end{subfigure}
    \hfill
    \begin{subfigure}[b]{0.49\columnwidth}
        \centering
        \includegraphics[width=\linewidth]{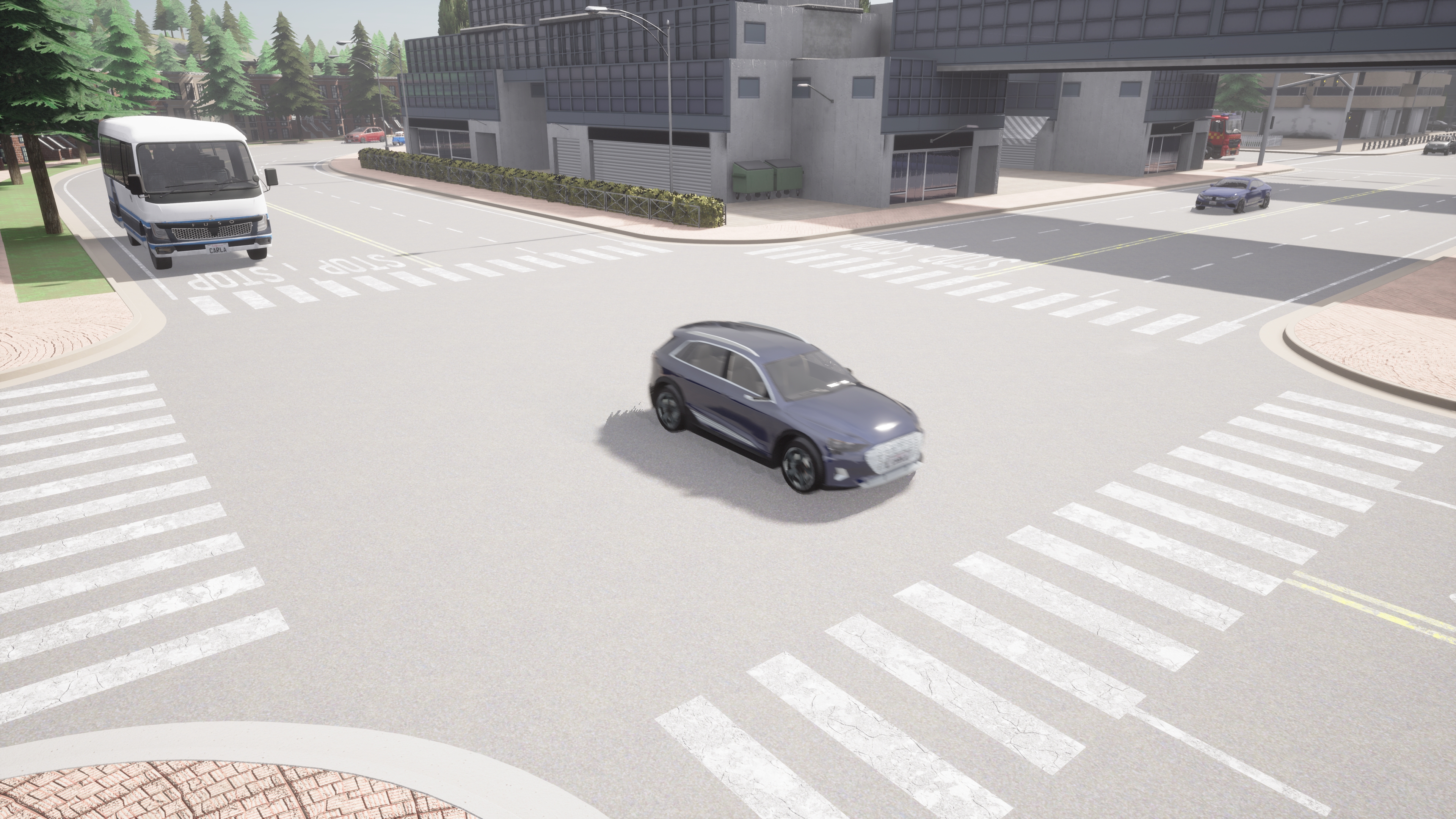}
        \caption{
            Sample frame generated using Carla - We calibrate the camera positions to mimic real videos.
        }
        \label{fig:eval-sample-images-carla}
    \end{subfigure}
    \caption{ 
    	Qualitative comparison of real and simulated data
     }
     \label{fig:eval-sample-images}
\end{figure}

\PP{City-scale video data generation}
We generate a large-scale city traffic video dataset using the Carla Simulator~\cite{carla}.
Carla is an open-source autonomous driving simulator built on top of Unreal Engine~\cite{unreal}.
% %
% It uses the OpenDRIVE standard~\cite{opendrive} to define roads and urban settings such as traffic signals.
%
It provides features such as realistic driving agents, synchronized simulation, and the ability to spawn cameras at arbitrary locations.
%
% Carla provides the ability to spawn cameras at arbitrary locations in the world.
%
% This allows us to generate video feeds that closely resemble real-world videos.
%
\cref{fig:eval-sample-images} visually compares frames from the Cityflow and Carla datasets.
Notice how we can calibrate the camera position to simulate real-world camera deployment.
Additionally, these cameras can synchronously capture the world at a specified frame rate for an arbitrarily long time.

\begin{figure}
    \centering
    \begin{subfigure}[b]{0.45\columnwidth}
        \centering
        \includegraphics[width=\linewidth]{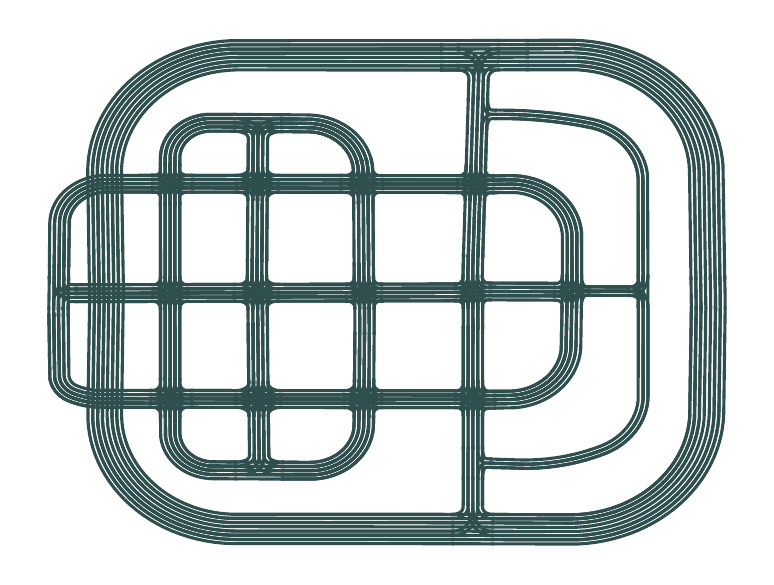}
           \caption{Road network of the Town05 map in Carla - cameras are deployed at the intersections in the road network}
           \label{fig:eval-sample-road-network}
    \end{subfigure}
    \hfill
    \begin{subfigure}[b]{0.45\columnwidth}
        \centering
        \includegraphics[width=\linewidth]{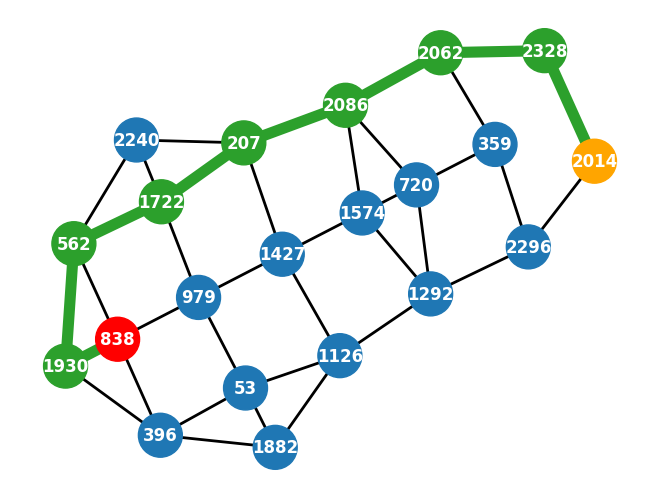}
        \caption{
           Camera graph generated automatically
     from the road network, with an annotated sample vehicle track
        }
        \label{fig:eval-sample-camera-trajectory}
    \end{subfigure}
    \caption{ 
    	Carla maps and corresponding camera networks
     }
     \label{fig:eval-carla-graph}
\end{figure}

% Specifically, we plan to build a framework that automatically spawns vehicles and cameras over a large (\eg 100 km$^2$) map and continuously generates vehicle tracking data for an arbitrarily long time.
%
Carla provides multiple maps with a varying number of road networks and intersections.
\cref{fig:eval-sample-road-network} shows an example map provided by Carla.
Using the Carla API, we can automatically spawn cameras at each intersection in the road network with minimal calibration.
We then automatically generate the camera network graph as shown in~\cref{fig:eval-sample-camera-trajectory} to use as input to \sys.
Carla offers the capability to incorporate custom road networks and world elements, enabling the simulation of extensive camera networks that adhere to real-world geographical limitations. 

Most importantly, Carla controls the spawning and driving behaviors of the vehicles.
We use CARLA Agent scripts~\cite{carlaagents} to simulate individual vehicle driving behavior.
Specifically, we set arbitrary spawn and destination points for each vehicle and then use local and global route planning algorithms to navigate from the spawn point to the destination point.
The vehicles navigate via the shortest path from source to destination while interacting with other vehicles, following traffic lights, signs, and speed limits.

\begin{figure}
    \centering
    \begin{subfigure}[b]{0.38\columnwidth}
        \centering
        \includegraphics[width=\linewidth]{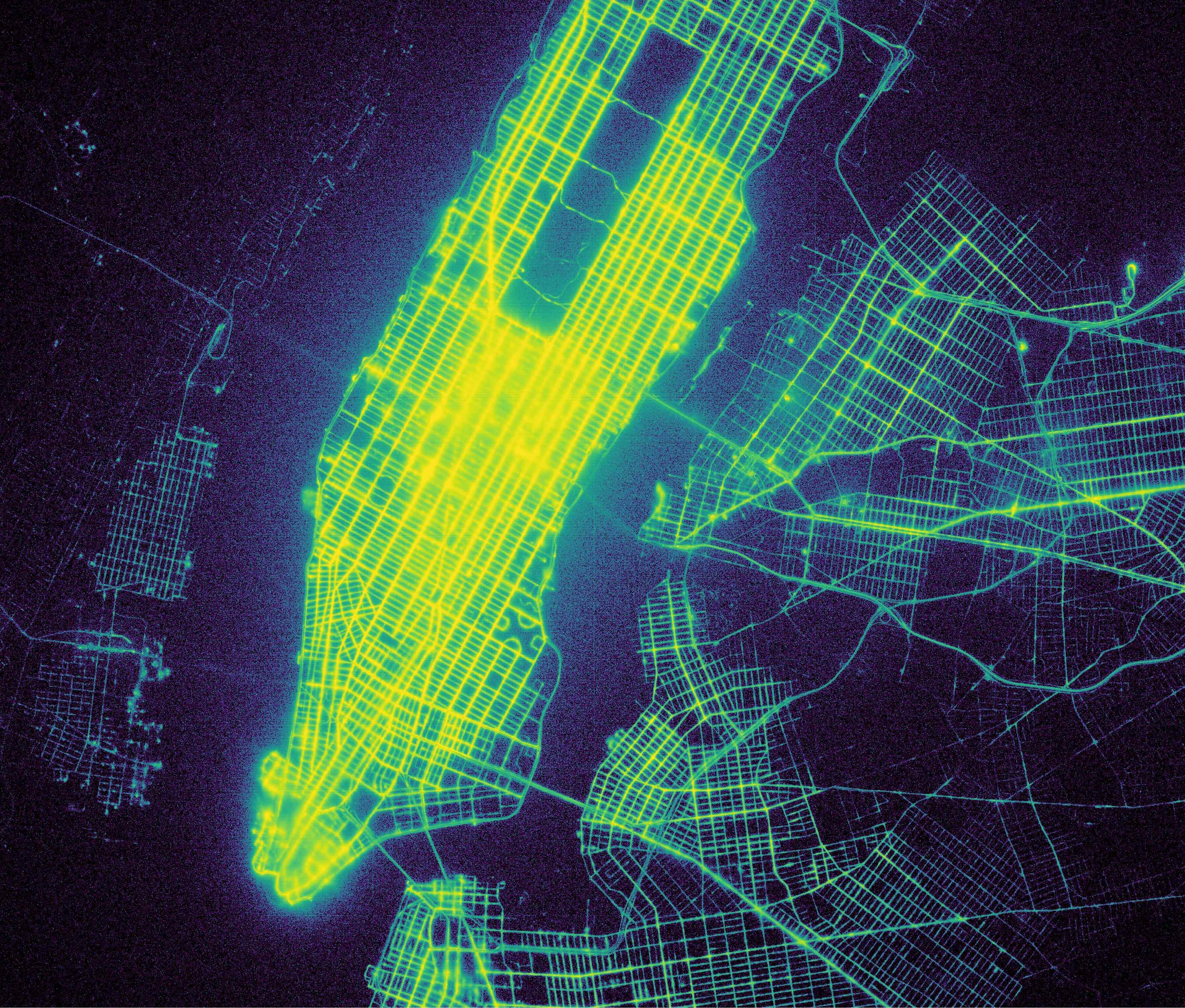}
           \caption{
                 Heatmap of pickup locations in NYC Taxi dataset.
            }
           \label{fig:nyc-taxi-pickups}
    \end{subfigure}
    \hfill
    \begin{subfigure}[b]{0.55\columnwidth}
        \centering
        \includegraphics[width=\linewidth]{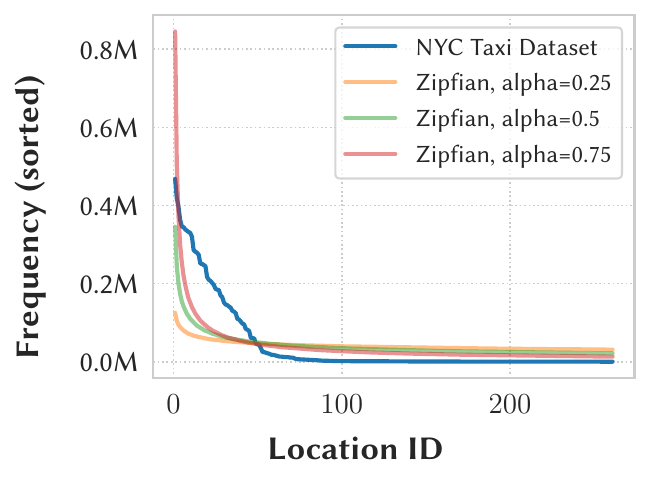}
        \caption{
            The pickup/dropoff locations loosely follow a Zipfian distribution.
        }
        \label{fig:zipfian-distribution-simulated}
    \end{subfigure}
    \caption{ 
    	Source/destination point distribution - We generate trajectory data by sampling the pickup and dropoff locations from a Zipf distribution with varying skew factors.
     }
     \label{fig:traffic-distribution}
\end{figure}

To simulate real-world traffic, we observe that city-scale traffic scenarios often exhibit hotspots, which are locations with a higher likelihood of being either the source or destination points.
For example, a few areas downtown and a few residential neighborhoods in a city are more likely to be source and destination points for the vehicle tracks.
We test this hypothesis on the real-world NYC Taxi dataset~\cite{nyc-taxi}.
\cref{fig:nyc-taxi-pickups} shows the heatmap of pickup locations for 10 million trips over a 1 month period.
As seen in~\cref{fig:zipfian-distribution-simulated}, the data loosely follows a Zipfian distribution with a few points accounting for most pickup locations.
So, we use a Zipfian distribution with different skew factors to generate the traffic patterns in our dataset.
Specifically, we sample the source and destination spawn points for each vehicle from the Zipfian distribution independently and navigate the vehicle through the map using the shortest path.
% \squishitemize
% \item Consider intersections as graph nodes
% \item Random vehicle spawning and constant vehicle number
% \item Zipfian distribution for vehicle start and end hotspots
% \item Normal driving behavior, but it can be changed to aggressive or cautious
% \item Reference to~\cref{fig:eval-carla-graph}.
% \squishend

\PP{Training data generation}
%
%\rev{
%
As discussed in~\autoref{sec:opt:recurrent-network}, we train the recurrent network using historical ground truth trajectories from a given camera network.
For the synthetic benchmark, the training data is generated by running the Carla simulator in a \textit{no-rendering mode}.
This mode produces ground truth traces 2$\times$ faster than visual data by simulating the whole environment without graphical rendering.
Our experiments show that 2000 ground truth traces suffice to train a highly accurate camera prediction model for all queries with a similar distribution.
The cost of generating 2K traces in the no-rendering mode is only 20$\times$ the per-query inference cost.
%
%}
%
\section{Evaluation}
\label{sec:evaluation}
% \notice{EVALUATION -- 
% List the questions being answered in the evaluation.
% }
%
In our evaluation, we illustrate that:
\squishitemize
\item \sys executes \reid queries 3.9$\times$ faster on average than the SoTA system \spatula, and 
4.7$\times$ faster on average than \dfs (\autoref{sec:eval-endtoend}).
\item \sys outperforms all the \reid baselines across various data skew levels and achieves close-to-\oracle performance in highly skewed distributions (\autoref{sec:eval-endtoend-distribution}).
\item \sys's recurrent network-based camera prediction module achieves 25\% higher accuracy than \spatula's localized prediction module (\autoref{sec:eval-camera-prediction-model}), and the performance gap widens as the camera networks grow larger (\autoref{sec:eval-endtoend-size}).
\item \sys incurs negligible runtime overhead compared to expensive vision operators, and \sys incurs a low per-query training cost (\autoref{sec:eval-cost-breakdown}).
\squishend

\begin{table*}[t]
    \small
    \centering
    \begin{adjustbox}{width=0.95\linewidth,center}
    \begin{tabular}{cccccccccc}
    \toprule
    %\multicolumn{3}{c|}{\textbf{\configuration{s}}} &  & \\\midrule
    \textbf{Topology} & \textbf{Data type}  & \textbf{\# Cameras} & \textbf{(Avg,  Max)} & \textbf{Total} & \textbf{Avg. \# objects} & \textbf{Avg. Trajectory} &\textbf{\# Training} \\
    &  &  & \textbf{Degree} & \textbf{Duration} & \textbf{per frame} & \textbf{Length (\# cameras)} & \textbf{Trajectories} \\ \midrule
    % \textbf{Dataset} & \textbf{No. of Classes} & \textbf{Train Size} &
    % \textbf{Test Size} & \textbf{Action Percent} \\\midrule
    \townfive & Synthetic Videos & 21 & (3.5, 4) & 10.4 hrs & 0.9 & 6.6 & 2298 \\
    \townseven & Synthetic Videos & 20 & (3.2, 4) & 16.0 hrs & 1.4 & 8.2 & 2104 \\
    \porto & GPS traces & 200 & (7.1, 8) & - & - & 13.0 & 25000 \\
    \beijing & GPS traces & 200 & (7.1, 8) & - & - & 9.9 & 7091 \\
    \bottomrule
    \end{tabular}
    \end{adjustbox}
    \caption{
        \textbf{Dataset Characteristics} -- 
        We report these key characteristics of the datasets we evaluate \sys on:
        (1) no.of cameras in the network,
        (2) avg. degree of the graph,
        (3) total no.of frames and avg. no.of objects per frame, and
        (4) average trajectory length and no.of training trajectories.
        }
    \label{tab:dataset-stats}
\end{table*}

\subsection{Experimental setup}
\label{sec:eval-setup}

\PP{Evaluation Settings}
We evaluate \sys on four network topologies from three datasets.
\cref{tab:dataset-stats} provides a summary of the key dataset statistics for the 4 topologies.
\squishitemize
\item \textbf{Carla-based synthetic scenarios:} We use the Carla-based synthetic scenarios generated using
the benchmark described in~\autoref{sec:eval-reid-benchmark} for the first two topologies.
\item \textbf{\porto:} For our third topology, we use the city of Porto, Portugal, and the Porto GPS trajectories dataset~\cite{porto}.
We designate 200 intersections in the Porto road network
as cameras locations in the graph.
The dataset contains 1.7 million GPS trajectories from 442 taxis over 12 months.
We use a subset of 25,000 trajectories for training and evaluation.
To simulate camera footage, we assume a camera deployed at each intersection captures a 200 square meter area at 10 fps.
Since we lack access to the full vehicle occupancy at intersections, we assume the same average occupancy as the Cityflow dataset~\cite{cityflow}.
%
% \PC{Mention not having access to all vehicle information at any given intersection? Cannot run the PP algorithm.}
%
\item \textbf{\beijing:} For the fourth topology, we utilize the Geolife GPS trajectories dataset collected in Beijing, China~\cite{beijing}.
This dataset includes 17,621 GPS trajectories annotated with 8 different transportation modes (walk, car, bus \etc) over 5 years.
We use trajectories from 4 modes - walk, car, taxi, and bus - for training and evaluation.
We use the same methodology as the \porto dataset to generate the camera graph and simulate the traffic footage.
\squishend
% We use two scenarios from the AI-city dataset as the first two topologies.
%

\PP{Baselines}
We compare \sys against five baselines.
\squishitemize
\item \naive algorithm uses an object detector and a \reid model (\autoref{fig:reid-pipeline}) on each frame of every
camera to detect the presence of the query object, terminating the search in a camera once the object is found~\cite{mtmcbaseline}.
%
% State-of-the-art computer vision frameworks~ use \naive, thus prioritizing \reid accuracy over efficiency.
%
\item \pp leverages lightweight proxy models, as used in prior
VDBMSs~\cite{noscope, pp, blazeit} 
to quickly filter irrelevant frames from the video.
Specifically, we use \pp to remove frames that do not contain any objects of the query class.
\item \dfs traverses the camera network starting from the source camera, randomly selecting and processing one neighboring camera at a time (\cref{fig:intro-graph-search}).
%
% \dfs also uses temporal priors to start the search from the most likely frame.
%
% So, \dfs uses static spatio-temporal filtering to accelerate \reid queries.
%
\item \spatula~\cite{spatula} utilizes localized camera history (\autoref{sec:opt:local-correlations}) to predict the
most likely next camera and start frame, as shown in~\cref{fig:intro-spatula}.
Since we operate in the 100\% recall setting, we do not leverage the replay search feature of \spatula and instead perform an exhaustive search.
\item \sys uses a combination of recurrent network-based camera prediction (\autoref{sec:opt:camera-prediction}) and probabilistic adaptive search model (\autoref{sec:opt:probabilistic-execution}) to process \reid queries.
\item \oracle is the ideal system that has access to the ground truth information
and hence chooses the best camera and frame to start the search from at each step (\autoref{sec:opt:problem-formulation}).
\squishend

\PP{Incremental Search Optimization}
Disabling the incremental search optimization leads to prohibitively longer queries with \dfs and \spatula (100$\times$ slower). 
Without this optimization, the systems perform an exhaustive search over all the frames from irrelevant cameras.
To better understand the utility of other optimizations in \sys (\ie the camera prediction module and the probability update algorithm), 
we enable the incremental search optimization in all three baselines (\dfs, \spatula, and \sys) across all the experiments.
% \PC{Add key takeaways at the end of each experiment}

\begin{figure}[t]
    \centering
    \hspace{2em}\includegraphics[width=0.7\linewidth]{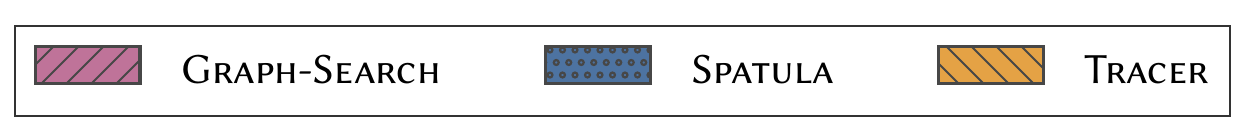}
    \includegraphics[width=\linewidth]{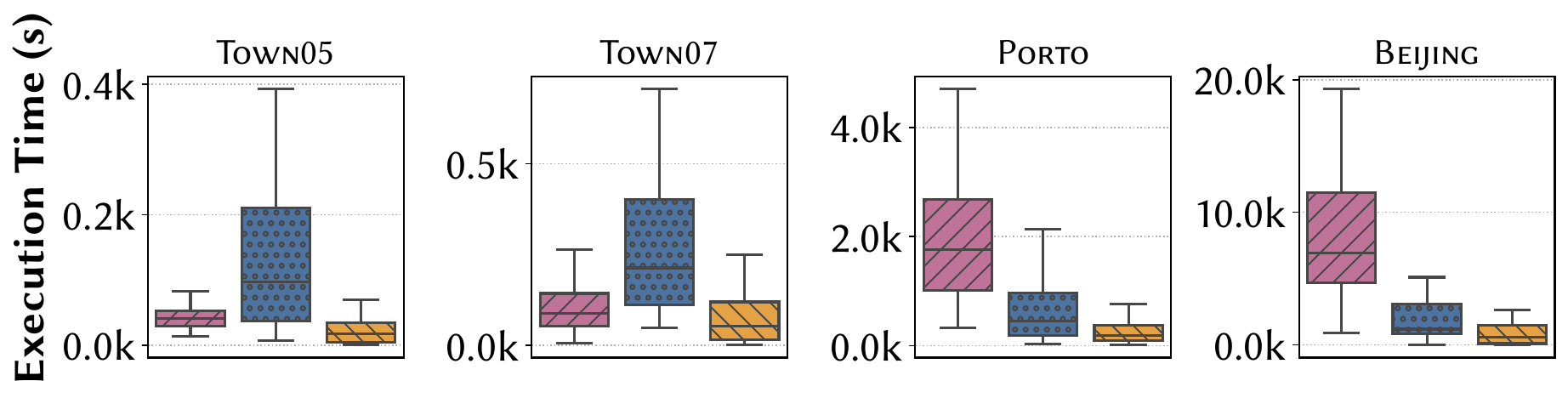}
    \caption{
     \textbf{End-to-end Performance Comparison --}
     The end-to-end execution time of the baselines and
     \sys over 4 topologies. 
     Note that the y-axis range is different for each query. 
     }
    \label{fig:eval-endtoend}
\end{figure}

\subsection{End-to-end Comparison}
\label{sec:eval-endtoend}
We first compare the end-to-end execution time of \sys
against the best-performing baselines, \dfs and \spatula.
As mentioned in~\autoref{sec:system-overview}, we process frames from each camera exhaustively to achieve 100\% recall for each baseline.
The throughput is averaged over 50 \reid queries, with each query executed 20 times
to account for variance from the stochastic neighboring camera and start frame selection.
The results comparing \sys to \dfs and \spatula are presented in~\cref{fig:eval-endtoend}.

\PP{\sys vs. \dfs} \sys executes \reid queries 4.7$\times$ faster than \dfs on average 
across all 200 queries spanning 4 topologies.
\sys outperforms \dfs by 7.8$\times$ on real-world \porto and \beijing topologies.
The higher average degree in these topologies (\autoref{tab:dataset-stats}) entails more adjacent cameras to process.
So, \sys benefits from the probability update algorithm that utilizes the high-confidence scores from the camera prediction module to greedily exploit frame windows from the correct cameras.
In contrast, \dfs's random exploration strategy incurs significant overhead before reaching the desired frame in the correct camera.
However, for synthetic \townfive and \townseven topologies with a lower average degree, \dfs performs reasonably well as its random exploration quickly explores the few neighbors.
So, \sys is only 1.7$\times$ faster than \dfs for these simpler topologies.

\PP{\sys vs. \spatula} \sys outperforms \spatula by 3.9$\times$ on average across all 200 queries.
This significant gain stems from \sys's recurrent network effectively capturing complex long-term cross-camera correlations, unlike \spatula's reliance on less accurate local correlations.
Since \spatula uses static probability scores during all sampling rounds, its weaker prediction model can cause exhaustive yet incorrect exploration of neighboring cameras.
Interestingly, \sys achieves a higher average speedup over \spatula on synthetic datasets compared to real-world datasets.
As seen in~\cref{fig:eval-endtoend}, \spatula exhibits high variance on synthetic queries.
This variance occurs because \spatula is sensitive to outliers in synthetic trajectories.
Synthetic datasets lack local context, only using a global probability distribution for source and destination cameras (\autoref{sec:eval-reid-benchmark}).
Thus, \spatula's local prediction model mispredicts outliers with high-confidence,  causing the high variance.
This highlights opportunities to improve synthetic data generation methods to better reflect real-world distributions.

\PP{\spatula vs. \dfs} \spatula outperforms \dfs on real-world datasets but is slower on synthetic datasets.
As discussed earlier, \spatula exhaustively processes incorrect neighboring cameras in synthetic datasets due to its inability to handle outliers.
In contrast, \dfs randomly and uniformly explores all the cameras, handling outliers more efficiently.
However, \dfs struggles on real-world datasets with high degree as it must uniformly explore several frame windows in neighboring cameras to arrive at the correct camera and frame.

%
% This is unexpected since \spatula uses the localized camera history to predict the nearby as opposed to the random exploration strategy in \dfs.
%
% In contrast, the random exploration strategy in \dfs is more effective
% in distributing the exploration effort across all the neighboring cameras, thereby resulting in a lower execution time.
%
% \PC{high degree in \porto and \beijing means overexploration even in \dfs.
% Rewrite to emphasize a balance between exploration and exploitation.}

\begin{figure}[t!]
    \centering
    \hspace{2.5em}\includegraphics[width=0.6\linewidth]{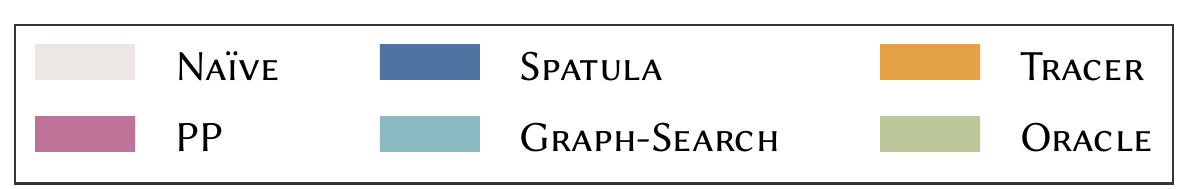}
    \includegraphics[width=0.7\linewidth]{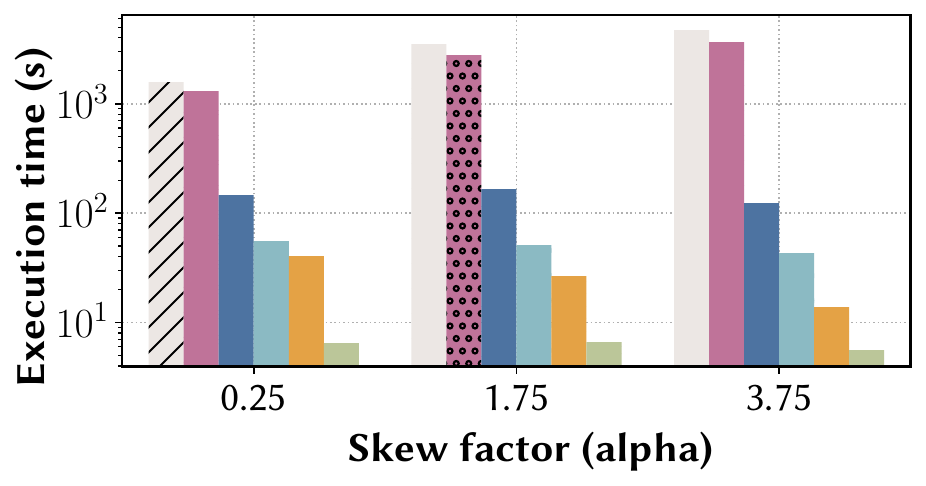}
    \caption{
     \textbf{Varying data skew --}
     The impact of data skew on the end-to-end performance of \sys and the other baselines in \townfive.
     }
    \label{fig:eval-varying-data-skew}
\end{figure}

\subsection{Impact of data distribution}
\label{sec:eval-endtoend-distribution}
In this experiment, we evaluate the impact of data distribution on the performance of \sys.
Specifically, we compare the performance of \sys against all the five baselines with changing
data skew on the \townfive topology.
Recall that data skew controls the complexity of long-term cross-camera correlations
in synthetic data (\autoref{sec:eval-reid-benchmark}).
A higher skew level generates vehicles trajectories with strong
long-term cross-camera correlations.
%
% Note that the data is generated from scratch for each skew level, and hence contains different
% vehicle trajectories and densities.
%
The results are shown in~\cref{fig:eval-varying-data-skew}.
Note that the y-axis is on a log scale to highlight the differences in the execution time.

\PP{\naive and \pp} \naive and \pp are the worst-performing methods across all data skews.
This is expected since they do not use information about the camera network topology
and hence cannot leverage the long-term cross-camera correlations.
Additionally, \pp only slightly outperforms \naive since the data has 
a high occupancy rate (\cref{tab:dataset-stats}).
%
% \PC{Reason for increase in execution time with increasing skew} The performance of...
%

\PP{\dfs and \spatula} \dfs and \spatula outperform \naive and \pp across all data skews as expected, since they leverage the network topology for camera prediction.
\dfs and \spatula perform marginally better at higher skew since vehicle trajectories are more predictable.
As seen in \autoref{sec:eval-endtoend}, \dfs performs better than \spatula at all skew levels due to the low average degree of synthetic datasets.

\PP{\sys} \sys is the best-performing learned method across all data skews.
The relative performance of \sys over the baselines increases with increasing
data skew.
This can be attributed to the fact that \sys's probabilistic search model benefits from the high confidence probability outputs of the recurrent network at high data skew.
Finally, we note that \sys's performance is comparable to \dfs on low skew data.
At low skew, the trajectories are not predictable, so the recurrent network in \sys
does not provide a significant advantage over random exploration.

\PP{\oracle} We also show the performance of \oracle at different data skews.
\oracle can predict the next camera and frame accurately as it has access to the ground truth (\autoref{sec:opt:problem-formulation}).
The gap between \oracle and \sys reduces with increasing data skew as the learned model in \sys
becomes more accurate.
At the highest skew level, the performance of \sys is only 2.5$\times$ lower than \oracle.

\subsection{Impact of camera prediction model}
\label{sec:eval-camera-prediction-model}
\begin{figure}[t!]
    \centering
    \hspace{2em}\includegraphics[width=0.7\linewidth]{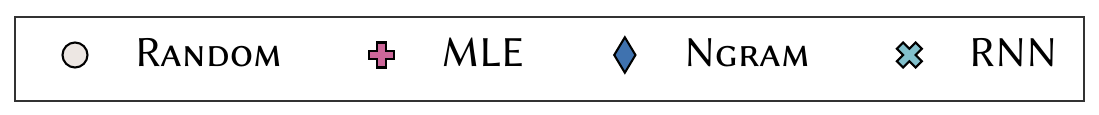}
    \includegraphics[width=\linewidth]{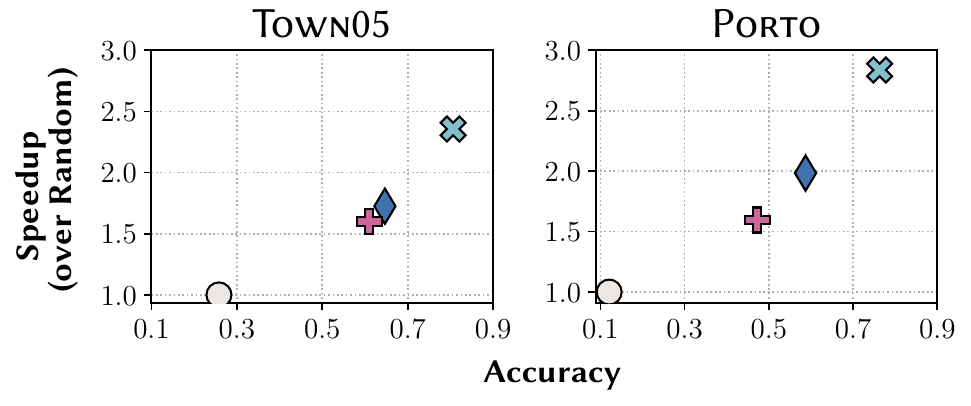}
    \caption{
     \textbf{Camera prediction models --}
     Speedup and accuracy of different camera prediction models
     in \sys over synthetic and real-world datasets.
     }
    \label{fig:eval-spatial-filtering}
\end{figure}

In this experiment, we evaluate the effectiveness of various models for camera prediction (\autoref{sec:opt:camera-prediction}).
We report the (1) speedup achieved by the models compared to the random traversal algorithm and (2) the camera prediction accuracy of the models.
The results are shown in~\cref{fig:eval-spatial-filtering}.
On aggregate, we observe that the accuracy and speedup of the models vary between real-world and synthetic datasets.
The models achieve a higher accuracy on synthetic data since the trajectories are more predictable.
However, this accuracy does not translate to a proportionally high speedup since the average degree
of the synthetic topologies is low, so incorrect camera predictions are not penalized as much.
%
% \PC{Explicitly state that the topology size in synthetic data can be controlled but needs city planners to construct openDrive maps.}
%

\PP{\ngram models}
%
% We utilize the \ngram model with best-performing length $n$ in this experiment.
%
\ngram model achieves lower accuracy compared to \rnn.
We observe that the \ngram model fails to capture the long-term cross-camera correlations in the data.
Specifically, we notice that the accuracy of the model drops significantly at larger $n$ ($> 3$).
We identify two reasons for this --
(1) the number of possible trajectories grows exponentially with $n$, increasing the complexity of the learning problem significantly
(2) In the training data, long trajectories (larger $n$) are scarce.
In contrast, the \rnn model automatically learns the necessary context from a given trajectory, and so, significantly outperforms \ngram.
Finally, \ngram outperforms the frequency estimation method \mle.
This demonstrates that considering context from multiple previous cameras improves next-camera prediction accuracy.
%
% \PC{Why is accuracy of \ngram low in synthetic data?}
%
% \PC{Have a discussion earlier on lower training data at much higher trajectory lengths, while larger trajectories have a larger impact on performance.}

\subsection{Size of the camera network}
\label{sec:eval-endtoend-size}
\begin{figure}[t!]
    \centering
    \hspace{2.5em}\includegraphics[width=0.6\linewidth]{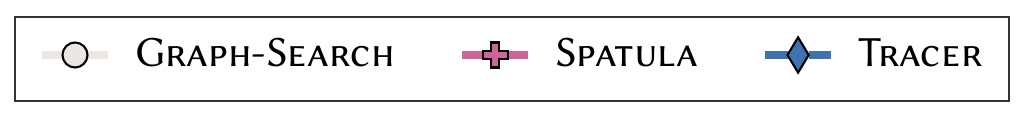}
    \includegraphics[width=0.8\linewidth]{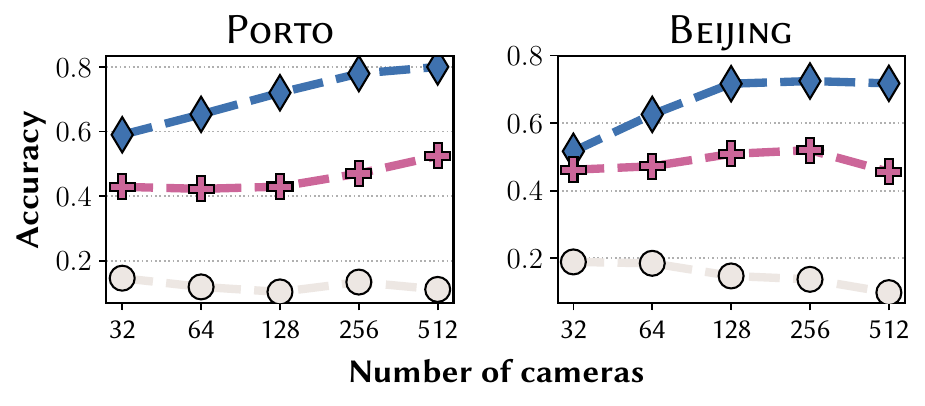}
    \caption{
     \textbf{Changing the camera network size --}
     Impact of camera network size on the accuracy of the learned model in \sys and the other baselines in \porto and \beijing.
     }
    \label{fig:eval-map-size-variation}
\end{figure}

In this experiment, we evaluate the impact of camera network topology size on the accuracy
of the camera prediction models in \sys.
Specifically, we compare the performance of \sys against \dfs and \spatula on real-world datasets with varying topology sizes (no.of cameras).
We fix the geographical area and increase the number of
cameras in the network by sampling more intersections from the map with the same network degree.
We report the accuracy of the models at different topology sizes in~\cref{fig:eval-map-size-variation}.

We first observe that the random exploration strategy in \dfs is unaffected by topology size as expected since the decisions are independent of the current trajectory and the network degree is unchanged.
%
% \PC{Need to emphasize earlier in eval that decisions depend on "the trajectory found so far".}
%
% \PC{Reason for the slightly decreasing accuracy trend of \dfs on \beijing?}
%
The accuracy of \sys's recurrent network increases with increasing topology size.
As the number of cameras in the network increases, more cameras capture the vehicle's movements within the same geographical area, resulting in more fine-grained and accurate vehicle trajectories.
This also explains the improving accuracy of \spatula's neighboring camera frequency
estimation model.
Importantly, the accuracy gap between \sys and \spatula widens for larger topologies as \sys is better at modeling long-term correlations.
%
% Finally, we note that the accuracy of all the models drops or plateaus at very large topology sizes,
% specifically for the \beijing dataset.
% %
% We find that the richness of available training data reduces as the number of cameras
% in the network becomes too large.
% %
% That is, the vehicle trajectories traversing several cameras is very less.
% %
% This reduction in training data is more pronounced in \beijing since the dataset only contains~17K trajectories
% as opposed to the 1.7M trajectories in \porto.
% %
% We find that the best-performing topology size with the available training data is around 200 cameras for both datasets.

\subsection{Execution speed analysis}
\label{sec:eval-cost-breakdown}

\PP{Cost breakdown}
\cref{fig:eval-cost-breakdown} shows the average cost breakdown per query for different operators in \sys
over 100 queries.
We observe that the \reid operator accounts for the highest cost in \sys, followed by the object detection models.
This is expected since the \reid model is invoked for every object detected in a given frame.
On the other hand, the camera prediction and frame prediction models have a negligible cost (100 ms)
since these models only require the current trajectories and not the raw video frames.
% \sys overhead is negligible, and the underlying vision operators account for most of the execution cost.

\PP{Training time}
Training time includes the time required to generate the training trajectories and train the models.
For synthetic datasets, generating training trajectories only requires 20$\times$ the per-query inference cost as discussed in~\autoref{sec:eval-reid-benchmark}.
For real-world deployment, the ground-truth trajectories can be generated offline by running the \reid pipeline on historical video feeds (\autoref{sec:opt:recurrent-network}).
Our experiments indicate that this offline process only requires $\approx$10$\times$ the per-query inference cost to generate a sufficient number of training trajectories.
% Training the models is a one-time cost for a given camera network.
% %
% So, once the camera topology is fixed, the models can be trained offline.
%
The \rnn model training cost is negligible - training an accurate \rnn model takes less than 5 minutes with 25K trajectories from the largest \porto topology.
%
% In contrast, the query execution time for 100 query trajectories is greater than
% 10 minutes for the same topology.
%
% Any new queries on this topology can leverage this trained model.
%
% \sys can manage the drift in traffic distributions~\cite{odin}
% through the periodic offline generation of ground-truth trajectories and retraining of the \rnn model.
% %
% Additionally, \sys can use inference trajectories from recent queries as ground-truth trajectories for retraining the models.
%

%

% \subsubsection{Parallel execution}

% \cref{fig:eval-parallel-execution} shows that \sys can be easily parallelized by executing multiple queries simultaneously.
% %
% Note that the performance depends on the number of GPU cores available to \sys.
% %
% \sys scales better than the baselines with the number of available GPUs since it significantly reduces the number of \reid model invocations.
% %
% \PC{TODO: does GPU run the queries in parallel or only schedules them? Should we show GPU utilization here?}

%% file: content/relatedwork.tex
\section{Related Work}
\label{sec:related-work}

\PP{Computer Vision}
%
% The object re-identification task can be broadly classified into two categories: person re-identification~\cite{personreidsurvey}
% and vehicle re-identification~\cite{vehiclereidsurvey}. 
%
% While techniques from one domain can be applied to the other~\cite{mtmcbaseline},
% \sys focuses on vehicle re-identification.
%
Multi-camera \reid is a well-studied problem in the field of Computer Vision~\cite{mtmcbaseline, mtmc2}.
Both traditional~\cite{multicamerasurvey, zapletal2016vehicle} and deep learning-based techniques~\cite{reidbaseline, mtmcbaseline, mtmc2} have been proposed for \reid.
These techniques primarily focus on improving the accuracy of the re-identification task, while \sys focuses on reducing the computational cost.

\begin{figure}[t!]
    \centering
    \includegraphics[width=0.85\linewidth]{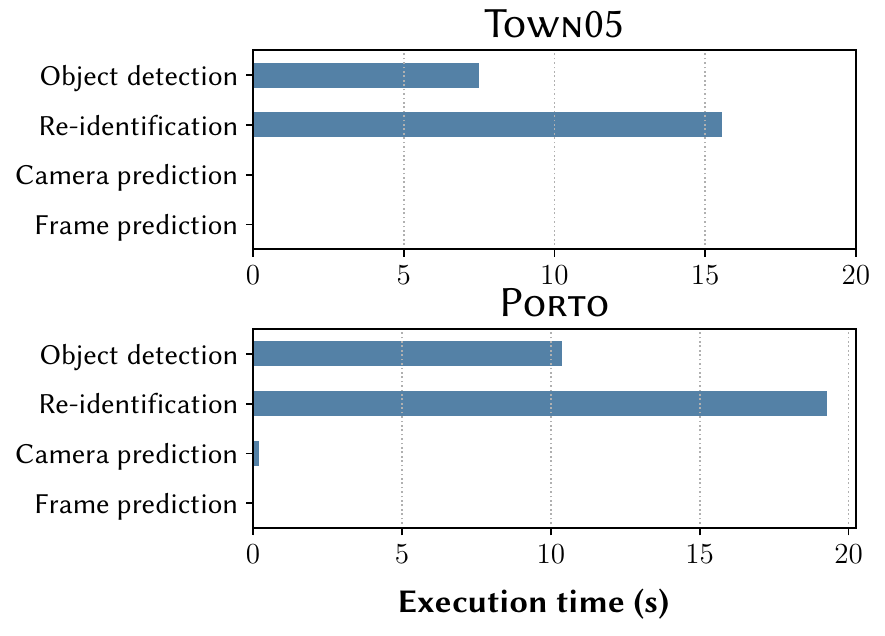}
    \caption{
     \textbf{Cost breakdown --}
     Cost breakdown of different operators involved in processing \reid queries in \sys.
     }
    \label{fig:eval-cost-breakdown}
\end{figure}

\PP{Video Data Management Systems}
Modern VDBMSs~\cite{blazeit, eva, viva, vocal, miris} primarily focus on the efficient retrieval of: 
(1) objects~\cite{pp, noscope, blazeit, eva},
(2) actions~\cite{zeus, equivocal, videomonitoring} and
(3) object tracks~\cite{miris, otif}.
These systems focus on a single-camera setting and do not consider object relationships across cameras.
In contrast, \sys focuses on efficiently processing \reid queries - it re-identifies \textit{specific object instances} across a camera network.

\PP{Cross-camera analytics}
Cross-camera analytics focuses on efficiently processing video queries that span multiple cameras.
Prior work in this area has mainly optimized online video analytics in the edge-computing
setting~\cite{spatula, crossroi, convince, adaptivecrosscam}.
Chameleon~\cite{chameleon} and Spatula~\cite{spatula} leverage localized spatio-temporal cross-camera correlations to reduce profiling cost and latency of live video analytics.
In contrast, \sys focuses on the offline setting and uses a novel probabilistic adaptive
query processing framework to reduce \reid query cost.
%
% While MFG~\cite{adaptivecrosscam} employs an adaptive video analytics framework,
% its primary emphasis lies in adapting to the evolving cross-camera correlations over time.
% %
% In contrast, \sys focuses on the adaptive and accurate selection of the most
% cost-efficient cameras to process, achieved through modeling long-term cross-camera correlations.
% Chameleon~\cite{chameleon}  spatula~\cite{spatula} CrossROI~\cite{crossroi} \cite{adaptivecrosscam} Convince~\cite{convince}

\PP{Adaptive Query Processing and Sensor Networks}
Adaptive query processing has been extensively explored to dynamically reorder query plans and optimize efficiency~\cite{eddies, aqpsurvey}.
%
% Traditionally, adaptive query processing approaches have focused on dynamically reordering query plans to improve query processing efficiency~\cite{}.
%
It has also been applied to sensor networks to adapt to resource changes and optimize data acquisition~\cite{tinydb, modeldriven, exploitingcorrelated}.
% \cite{modeldriven} \cite{exploitingcorrelated} \cite{modelbasedapproximate} ~\cite{usingprobabilistic} 
%
\sys draws inspiration from these works to effectively and adaptively select the most cost-efficient cameras at each time step.
Additionally, it incorporates a probabilistic adaptive search model, inspired by~\cite{usingprobabilistic}, to optimize \reid queries.

\PP{Spatio-temporal data mining}
%
%\rev{
%
In spatio-temporal data mining, recurrent neural networks have often been used for spatio-temporal location prediction.
These methods apply large recurrent network models to dense location data, such as GPS traces, to capture spatial-temporal context and improve the accuracy of next location prediction~\cite{liu2016predicting, al2016stf, xiao2020vehicle, altaf2018spatio}.
In contrast, \sys applies recurrent networks to sparse camera trajectories obtained from video analytics primitives.
It focuses on improving \reid query efficiency and utilizes the recurrent network to optimize the adaptive search process.
%
% \PC{Our goal is also different, accuracy vs }
%}
%

%% file: content/conclusion.tex
\section{Conclusion}
\label{sec:conclusion}

% \notice{In the conclusion, you can assume that the reader has gone over the
% paper. So, you can refer to specific details of your solution.}
%
Efficiently re-identifying and tracking objects across a network of cameras is an important vision problem in practice.
In this paper, we presented \sys, a novel VDBMS tailored for answering such \reid queries using two key optimizations.
First, it trains a recurrent network to model the long-term spatial correlations, thereby improving the camera prediction accuracy.
Second, it utilizes a probabilistic adaptive search model that processes camera feeds in incremental windows, and adaptively adjusts the probabilities based on an exploration-exploitation trade-off.
Evaluation on three diverse datasets shows that \sys accelerates \reid query processing by~3.9$\times$ on average compared to \spatula.
%